 \documentclass[preprint2]{aastex}

\usepackage[bookmarks=false,         
     pdfnewwindow=true,      
     colorlinks=true,    
     linkcolor=cyan,     
     citecolor=cyan,     
     filecolor=cyan,  
     urlcolor=cyan,      
 ]{hyperref} 
\usepackage[utf8]{inputenc} 

\usepackage{breakurl}
\usepackage{color}

\def\heii{\ion{He}{2}~4686~\r{A} }
\def\hheii{\ion{He}{2}~4686~\r{A}}

\shorttitle{Exploratory Spectroscopy of Magnetic CVs Candidates}
\shortauthors{Oliveira et al.}

\begin{document}


\title{Exploratory Spectroscopy of Magnetic Cataclysmic Variables Candidates and Other Variable Objects}



\author{A. S. Oliveira \altaffilmark{}}
\affil{IP\&D, Universidade do Vale do Paraíba, 12244-000,
    São José dos Campos, SP, Brazil}
\email{alexandre@univap.br}

\author{C. V. Rodrigues\altaffilmark{}, D. Cieslinski\altaffilmark{}, F. J. Jablonski\altaffilmark{}}
\affil{Divisão de Astrofísica, Instituto Nacional de Pesquisas Espaciais, 12227-010, 
    São José dos Campos, SP, Brazil}

\author{K. M. G. Silva\altaffilmark{}}
\affil{Gemini Observatory, Casilla 603, La Serena, Chile}

\author{L. A. Almeida \altaffilmark{}}
\affil{Instituto de Astronomia, Geofísica e Ciências Atmosféricas, Universidade de São Paulo,
    05508-900, São Paulo, SP, Brazil}

\author{A. Rodr\'{\i}guez-Ardila \altaffilmark{}}
\affil{Laboratório Nacional de Astrofísica LNA/MCTI, 37504-364, Itajubá MG, Brazil}

\and    

\author{M. S. Palhares \altaffilmark{}}
\affil{IP\&D, Universidade do Vale do Paraíba, 12244-000,
    São José dos Campos, SP, Brazil}




\begin{abstract}

The increasing number of synoptic surveys made by small robotic telescopes, such as the photometric Catalina Real-Time Transient Survey (CRTS), represents a unique opportunity for the discovery of variable sources and improves the statistical samples of such classes of objects. Our goal is the discovery of magnetic Cataclysmic Variables (mCVs). These are rare objects, which probe interesting accretion scenarios controlled by the white dwarf magnetic field. In particular, improved statistics of mCVs would help to address open questions on their formation and evolution. We performed an optical spectroscopy survey to search for signatures of magnetic accretion in 45 variable objects selected mostly from the CRTS. In this sample we found 32 CVs, 22 being mCV candidates from which 13 are previously unreported as such. If the proposed classifications are confirmed, it would represent an increase of 4\% in the number of known polars and 12\% in the number of known IPs.
A fraction of our initial sample was classified as extragalactic sources or other types of variable stars by the inspection of the identification spectra. Despite the inherent complexity in identifying a source as a mCV, variability-based selection followed by spectroscopic snapshot observations has proved to be an efficient strategy for their discoveries, being a relatively inexpensive approach in terms of telescope time.

\end{abstract}


\section{INTRODUCTION} \label{sec:intro}

Cataclysmic Variable stars (CVs) are binary systems consisting of a white dwarf accreting mass from a less massive main sequence or a slightly evolved companion star via Roche lobe overflow \citep[e.g.,][]{warner1995}. For most systems, the secondary is a non-degenerate object, so the mass transfer leads to a decrease in the orbital period and demands a loss of orbital angular momentum that is usually attributed to two complementary mechanisms: magnetic braking of the secondary star, important in longer orbital period systems, and gravitational radiation losses that dominates in shorter period systems, in which the former mechanism is less efficient. 
See \citet{2011ApJS..194...28K} for a nice review of CV evolution.
Most commonly, the accretion proceeds via an extended accretion disk that surrounds the accreting white dwarf. The disk
is usually the dominant source of radiation in the binary system during periods of high accretion. The accretion disk may suffer recurrent episodes of instability known as Dwarf Nova (DN) outbursts, which increase the luminosity of the system by $2-6$ magnitudes in timescales as short as a day.

If the white dwarf has a strong enough magnetic field, however, the transferred ionized material connects to the magnetic field lines and accretes directly to the white dwarf surface, close to the magnetic poles, via an accretion column instead of a disk. These systems are known as magnetic Cataclysmic Variables (mCVs) and can be classified either as polars, whose magnetic fields are intense (B~$\sim 7-230$~MG), or as intermediate polars (IPs), in which B~$\sim 1-10$~MG. The distinctive characteristic of the polars is the synchronization of the white-dwarf spin and the orbital period, caused by magnetic locking of the primary and secondary stars \citep[see][for a review on polars]{1990SSRv...54..195C}. Matter falling through the accretion column emits cyclotron radiation, which is highly polarized, anisotropic and dominates the optical and infrared emission. Intermediate polars, on the other hand, are not synchronized systems, and the distinct spin, beat and orbital periods manifest in optical and X-rays. IPs usually present an accretion disk, truncated in its inner region by the magnetosphere of the white dwarf. 

A large number of CV systems has been discovered by surveys based on color selection, such as those identified through SDSS data  \citep[e.g.,][]{2011AJ....142..181S,2000AJ....120.1579Y}. Surveys based on variability are also prolific in finding new CVs and in the near future, the Large Synoptic Survey Telescope - LSST - is expected to strongly impact the statistics of CVs. One of these variability surveys is the Catalina Real-Time Transient Survey - CRTS \citep{2009ApJ...696..870D,2014MNRAS.441.1186D}. CRTS combines three distinct surveys - the Mount Lemmon Survey (MLS) and the Catalina Schmidt Survey (CSS) in the Northern Hemisphere and the Siding Spring Survey (SSS) in the South - covering an area of 30,000 deg$^2$ with a typical cadence of two weeks, reaching a depth of 19-21 magnitudes. The CRTS has already found thousands of CV candidates, which are rapidly announced via their website
\footnote{\url{http://crts.caltech.edu}} and are subject to follow-up investigations \citep[e.g.,][]{2014MNRAS.437..510C,2014MNRAS.441.1186D,2014MNRAS.443.3174B,2012AJ....144...81T,2016MNRAS.456.4441C}. As the prime criterion for the CRTS detecting transients is the brightening by 2 or more magnitudes, it turns out to be very effective in discovering non-magnetic CVs, as the follow-up works mentioned above have shown. Other variability surveys that found new CVs are the All Sky Automated Survey (ASAS; \citealt{2002AcA....52..397P}), the Palomar Transient Factory (PTF; \citealt{2014ASPC..490..389M,2015MNRAS.446..391L}) and the Optical Gravitational Lensing Experiment (OGLE; e.g.  \citealt{2013AcA....63..135M}). 

The primary goal of this paper is to increase the number of known mCVs systems, which is justified below. These systems are relevant for studying and understanding the physics of magnetically controlled accretion. This phenomenon is widespread in astronomy, from stellar formation processes to accretion on compact objects. But mCVs are relatively simple objects in which the magnetic accretion is observed practically free of other effects allowing a precise description of the process. The short-term variability of mCVs, spanning from minutes to hours, enables us to disentangle the system geometry with a relatively small set of observations. An important aspect of CV research that can be addressed by an improved statistics of members is the evolution of mCVs, specially the relationship between IPs and polars \citep[e.g.][]{2015SSRv..191..111F,2013MNRAS.432..570P}. For instance, it is not known if the surface magnetic field changes along the mCV evolution. mCVs, IPs in particular, are rare objects and the incompleteness and biases of the present samples are not clear. To determine the frequency distribution of properties as orbital period and primary magnetic field, it is necessary to study individual systems using time-resolved observations and proper modeling \citep[e.g.,][]{2009MNRAS.398..240C,2015MNRAS.451.4183S}. Polarimetry is particularly insightful for this, since cyclotron emission is a direct evidence of magnetic fields \citep[e.g.,][]{2006MNRAS.369.1972R}. In particular, our group is building a fast-camera tailored to CVs studies that will perform simultaneous photometry and polarimetry in 4 broad bands \citep{2012SPIE.8446E..26R}. The 3D modeling of the geometry of the magnetic accretion is an important tool to a realistic representation of mCVs. This kind of analysis can be performed using the Cyclops code, developed by us to study stellar magnetic accretion \citep{2009MNRAS.398..240C,2013MNRAS.432.1587S}.

Here we address the selection of mCVs candidates for follow-up studies. This paper should be understood as a first step in a broader and ongoing study of mCVs. We performed snapshot spectroscopy of a sample of selected objects. The most promising or peculiar objects will be subject of follow-up time-resolved observations to confirm and/or characterize the systems. Using this strategy, we optimize the use of telescope time. 

This paper is organized as follows. Section \ref{sec_selection} presents the general rules for the sample selection. Section \ref{sec_observations} describes the spectroscopic observations. In Section \ref{sec_time_analysis} we present the methods for the time series analysis of the CRTS photometric data. Section \ref{sec_classification_criteria} describes the criteria applied for the classification of the individual objects of the sample. The spectroscopic results and proposed classification for each object are shown in Section \ref{sec_classification}.  A summary and discussion of the main results are in Section 7.

\section{SAMPLE SELECTION}
\label{sec_selection}

In this section we describe the criteria employed to select the final sample of 45 variable objects. We then obtained exploratory spectra (Sec.~\ref{sec_observations}) aiming to identify spectral features that could be signatures of high-ionization mass accretion, typical of mCVs (see Sec.~\ref{sec_classification_criteria}).

Our sample is mainly composed by CRTS objects. We also included in the sample mCV candidates mentioned in the literature that have no published spectroscopic measurements. Previous reports of X-ray detections via VSNET\footnote{http://ooruri.kusastro.kyoto-u.ac.jp/mailman/listinfo/vsnet-alert} and ATel\footnote{http://www.astronomerstelegram.org/} alerts were also considered. The classification of CVs in general and mCVs in particular can be a difficult task and may demand different observational techniques (as discussed in Sec. \ref{sec_classification_criteria}). This justifies the inclusion of already known mCVs candidates in our sample. Besides, the objects should have suitable apparent magnitudes and positions in the sky for SOAR and OPD observations. No color criterion was applied to define the sample.

The selection of the CRTS objects was done by visual inspection of the light curves with at least one of the following characteristics:  

\renewcommand{\theenumi}{\roman{enumi}}
\begin{enumerate}

\item Intrinsic flux variability, which may indicate accretion processes. The flux dispersion should be larger than the observational error expected for the mean magnitude;

\item Transitions between two states of brightness, as seen in polars;

\item No or few outbursts, in order to minimize dwarf nova detections, although we cannot simply exclude objects with outbursts, since IPs can present such events.
 
\end{enumerate}

\section{SPECTROSCOPIC OBSERVATIONS AND DATA REDUCTION}
\label{sec_observations}

The identification spectra were obtained using the SOAR 4.1-m telescope at Cerro Pachón, Chile, and the Perkin-Elmer 1.6-m telescope at Observatório do Pico dos Dias (OPD-LNA/MCTI), located in Southeast Brazil. The RA ordered list of observed objects is shown in Table \ref{obs}. The identification of the objects from the CRTS is in the form TTTyymmdd:hhmmss$\pm$ddmmss, where TTT indicates one of the three dedicated telescopes of the survey -- CSS, SSS or MLS -- yymmdd is the discovery date of the transient and hhmmss$\pm$ddmmss are the target coordinates. In order to avoid confusion between transients discovered by the same telescope at the same date, and for simplicity, we will abbreviate these IDs as TTThhmm$\pm$dd. 

The SOAR Telescope was operated in service mode and the spectra were obtained using the Goodman High Throughput Spectrograph \citep{2004SPIE.5492..331C}, which employs Volume Phase Holographic  (VPH) gratings to maximize throughput, reaching down to the atmospheric cutoff at 3200 \r{A}. It is equipped with a Fairchild 4096$\times$4096 CCD with $15\times15$ micron pixel$^{-1}$ (0.15 $\arcsec$ pixel$^{-1}$). The spectrograph was set to operate with the 600 l mm$^{-1}$ grating, 1.68$\arcsec$ slit, and GG~385 blocking filter, yielding a spectral resolution of 7~\r{A} FWHM in the range $4350-7005$ \r{A}. 
Three exposures of each science target were obtained to remove cosmic rays. Cu-He-Ar lamp exposures were obtained for wavelength calibration, which resulted in typical 0.8 \r{A}, or about 45 km s$^{-1}$, RMS residuals. The [\ion{O}{1}] 5577, 6300 and 6363 \r{A} telluric  lines were used to assess the calibration accuracy. The slit was aligned to the parallactic angle to avoid light losses due to the atmospheric differential refraction. Bias images and quartz lamp calibration flats were taken to correct for the detector read-out noise  
and sensitivity.  Spectra of spectrophotometric standards \citep{1992PASP..104..533H} were used for flux calibration. 

The OPD-LNA observations were carried out using a Boller \& Chivens Cassegrain spectrograph on March and September, 2012. A 300 l mm$^{-1}$ grating, blazed at 6400 \r{A}, was used in these two observing runs, providing a spectral coverage in the range $4100-8530$ \r{A}. The slit aperture was set to 2.50$\arcsec$ in March and 2.00$\arcsec$ in September, giving a spectral resolution of 7~\r{A} and 6~\r{A} (FWHM), respectively. Two different thin, back-illuminated CCD detectors were used: an iKon-L936-BV (March) and an iKon-L936-BR-DD (September), both with 2048$\times$2048 pixels and 13.5$\times$13.5 micron pixel$^{-1}$. In the March run, a GG~385 blocking filter was used, while in September no filter was necessary. All targets were observed with integration times of 600 or 900 s. Multiple exposures (normally 2 or 3) were taken aiming to improve the S/N ratio and to minimize effects of cosmic rays in the images. He-Ar comparison lamp exposures were obtained after each target observation for wavelength calibration. For flux calibration, we observed the spectrophotometric standard stars HR 3454 and HR 9087 from \citet{1992PASP..104..533H}. In all nights, images of dome flats and bias were also obtained.

The data reduction, spectra extraction and wavelength, extinction and flux calibrations were performed using standard IRAF\footnote{IRAF is distributed by the National Optical Astronomy Observatories, which are operated by the Association of Universities for Research in Astronomy, Inc., under cooperative agreement with the National Science Foundation.} routines.


\begin{deluxetable}{llllclcl}
\tabletypesize{\scriptsize}
\tablecaption{List of observed targets. 
\label{obs}}
\tablewidth{0pt}
\tablehead{
\colhead{Object name} & \colhead{Abbreviation} & \colhead{RA (J2000)} & \colhead{Dec (J2000)} & \colhead{Date obs.} & \colhead{Telesc.} &
\colhead{Exp. time (s)} & \colhead{Type\tablenotemark{a}}
}
\startdata
CSS091009:010412-031341 & CSS0104-03  &   01:04:12 & -03:13:41  & 2012 Aug 25    & SOAR &  3600   &  D/IP        \\    
CSS091215:021311+002153 & CSS0213+00 &   02:13:11 & +00:21:53  & 2012 Sep 09    & SOAR &  8100   &  E        \\
MLS110213:022733+130617 & MLS0227+13  &   02:27:33 & +13:06:17  & 2012 Sep 09    & SOAR &  3600   &  P     \\
CSS071206:031525-014051 & CSS0315-01  &   03:15:25 & -01:40:51  & 2012 Nov 13    & SOAR &  8100   &  E          \\
CSS090922:032603+252534 & CSS0326+25  &   03:26:03 & +25:25:34  & 2012 Nov 20    & SOAR &  3600   &  D           \\
CSS091109:035759+102943 & CSS0357+10  &   03:57:59 & +10:29:43  & 2012 Nov 12    & SOAR & 8100    &  P       \\
MLS101203:045625+182634 & MLS0456+18  &   04:56:25 & +18:26:34  & 2012 Nov 13    & SOAR & 8100    &   P      \\
XMMSL1 J063045.9-603110 & XMM0630-60  &   06:30:45 & -60:31:13  & 2012 Nov 12 & SOAR &  8100     &   N        \\
MLS101226:072033+172437 & MLS0720+17  &   07:20:33 & +17:24:37  & 2012 Nov 12 & SOAR &   1440   &   P              \\
1RXS J072103.3-055854 & 1RXS0721-05   &   07:21:03 & -05:59:20 & 2012 May 01 & SOAR & 1440 & HA     \\   
MLS120127:085402+133633 & MLS0854+13  &   08:54:02 & +13:36:33  & 2012 Dec 15 & SOAR &  8100    &   P              \\
1RXS J100211.4-192534 & 1RXS1002-19 &   10:02:11 &  -19:25:36  & 2012 May 30 & SOAR &  360    &   P             \\
CSS120324:101217-182411 & CSS1012-18  &   10:12:17 &  -18:24:11  & 2012 May 30 & SOAR &  3600    &   D/IP             \\
SSS110504:101240-325831 & SSS1012-32  &   10:12:40 &  -32:58:31  & 2012 Apr 21 & SOAR &  3600    &   D              \\
CSS110225:112749-054234 & CSS1127-05  &   11:27:49 &  -05:42:34  & 2012 Mar 18 & SOAR &   1440   &   P              \\
CSS071218:124027-150558 & CSS1240-15  &   12:40:27 &  -15:05:58  & 2012 Mar 17 & SOAR &  3600    &   E              \\
MLS110329:125118-020208 & MLS1251-02  &   12:51:18 &  -02:02:08  & 2012 Apr 23 & SOAR &  8100    &   E              \\
SSS110724:135915-391452 & SSS1359-39  &   13:59:15 &  -39:14:52  & 2012 Apr 22 & SOAR &  1440    &   D/IP             \\
MLS110301:140203-090329 & MLS1402-09  &   14:02:03 &  -09:03:29  & 2012 Apr 23 & SOAR &  8100    &   E         \\ 
MLS100617:140447-152226 & MLS1404-15  &   14:04:47 &  -15:22:26  & 2012 Mar 17 & SOAR &  3600    &   E             \\
SSS100507:144833-401052 & SSS1448-40  &   14:48:33 &  -40:10:52  & 2012 Mar 26 & OPD  &  1800    &   PRG          \\
CSS100216:150354-220711 & CSS1503-22  &   15:03:54 &  -22:07:11  & 2012 Mar 16 & SOAR &  3600    &   P             \\
MLS110526:151937-130602 & MLS1519-13  &   15:19:37 &  -13:06:02  & 2012 Mar 18 & SOAR &  3600    &   RRL              \\
MLS110609:160907-104013 & MLS1609-10  &   16:09:07 &  -10:40:13  & 2012 Apr 22 & SOAR &  3600    &   P              \\
CSS080606:162322+121334 & CSS1623+12  &   16:23:22 &  +12:13:34  & 2012 Apr 22 & SOAR &  8100    &   D              \\
SSS100804:163911-235804 & SSS1639-23  &   16:39:11 &  -23:58:04  & 2012 Sep 03-04 & OPD  &  2700    &   PRG      \\
CSS110623:173517+154708 & CSS1735+15  &   17:35:17 &  +15:47:08  & 2012 Apr 23 & SOAR &  1440    &   D              \\
1RXS J174320.1-042953 & 1RXS1743-04 &   17:43:20 &  -04:29:57  & 2012 Mar 16 & SOAR & 720  & P    \\
1RXS J192926.6+202038 & 1RXS1929+20 &   19:29:28 &  +20:20:35  & 2012 Apr 23 & SOAR & 1440  & D    \\
SSS110625:194030-633056 & SSS1940-63  &   19:40:30 &  -63:30:56  & 2012 Sep 03 & OPD  & 2700     &   PRG            \\
SSS100805:194428-420209 & SSS1944-42  &   19:44:28 &  -42:02:09  & 2012 Apr 22 & SOAR & 3600     &   P              \\
SSS110526:195648-603430 & SSS1956-60  &   19:56:48 &  -60:34:30  & 2012 Apr 21 & SOAR &  720    &   P              \\
USNO-A2.0 0825-18396733 & USNO0825-18 &   20:31:38 &  -00:05:11   & 2012 May 20-Jun 25 & SOAR &   3600    &   P  \\
SSS110526:204247-604523 & SSS2042-60  &   20:42:47 &  -60:45:23  & 2012 Aug 10 & SOAR & 3600     &   HA               \\
MLS100706:204358-194257 & MLS2043-19  &   20:43:58 &  -19:42:57  & 2012 Aug 10 & SOAR & 540     &   D/IP             \\
MLS111021:204455-162230 & MLS2044-16  &   20:44:55 &  -16:22:30  & 2012 Sep 03 & OPD  & 2700     &   RRL              \\
MLS101102:205408-194027 & MLS2054-19  &   20:54:08 &  -19:40:27  & 2012 Aug 10 & SOAR &  1800    &   D/IP             \\
CTCV J2056-3014 & CTCV2056-30         &   20:56:52 &  -30:14:38  & 2012 Sep 04 & OPD  &  2700  & D/IP    \\ 
CSS110513:210846-035031 & CSS2108-03  &   21:08:46 &  -03:50:31  & 2012 May 20-Jun 25 & SOAR &  3600    &   D    \\
MLS100620:213227-150523 & MLS2132-15  &   21:32:27 &  -15:05:23  & 2012 Aug 10 & SOAR & 900     &   RRL              \\
CSS100624:220031+033431 & CSS2200+03  &   22:00:31 &  +03:34:31  & 2012 Aug 14 & SOAR & 3600     &   D/IP        \\
1RXS J222335.6+074515 & 1RXS2223+07 &   22:23:34 &  +07:45:19  & 2012 Aug 25 & SOAR & 8100 & D   \\
MLS100906:223034-042347 & MLS2230-04  &   22:30:34 & -04:23:47  & 2012 Aug 25 & SOAR & 1440    &  E              \\
SSS110720:224200-662512 & SSS2242-66  &   22:42:00 & -66:25:12  & 2012 Sep 03-04 & OPD  &  2700   & D/IP              \\
CSS111021:231909+331540 & CSS2319+33  &   23:19:09 & +33:15:40  & 2012 Sep 04 & OPD  & 2700    &  D/IP            \\
\enddata
\tablenotetext{a}{P: polar candidate, D: disk system, D/IP: IP candidate, HA: system with high-state accretion disk,  N: Nova, RRL: RR~Lyrae star, PRG: Mira or low-amplitude Pulsating Red Giants star, E: extragalactic source.  }
\end{deluxetable}

\section{TIME SERIES ANALYSIS OF CRTS DATA}
\label{sec_time_analysis}

We performed time series analyses of the CRTS photometric data for all Galactic objects to search for periodic signals. The sampling of the CRTS light curves has a typical minimum separation between measurements of $\Delta$t~=~0.01 days. This means that if the time series was equally spaced with this sampling, the Nyquist frequency would be F$_{Ny}$ = 1/ 2$\Delta$t = 50 d$^{-1}$. The total span of the observations is $\approx$3000 d, which implies (again for equally spaced sampling) a frequency step $\Delta$F = 1/T = 3.3 $\times$ 10$^{-4}$ d$^{-1}$. The number of frequencies to be examined for coherent signals is N = F$_{Ny}$/$\Delta$F $\approx$ 1.5 $\times$ 10$^5$. We choose to examine a grid of N = 10$^6$ frequencies giving some slack to allow for possible effects coming from the nonuniform sampling of the time series data. We are aware that sparse data, as CRTS data, can produce complex structure in the power spectrum due to lateral lobes and aliasing. However, the periods quoted in this article have always a false alarm probability (FAP) less than 0.01 (=1\%) - see below how it is quantified.

We used the Discrete Fourier Transform (DFT, \citealt{1975Ap&SS..36..137D,1976Ap&SS..39..447L}), the Lomb-Scargle (LS, \citealt{1976Ap&SS..39..447L,1982ApJ...263..835S}) including its Bayesian formalism implementation (BGLS, \citealt{2015A&A...573A.101M}), the String-Length (SL, \citealt{1983MNRAS.203..917D,2002A&A...386..763C}) and the Phase Dispersion Minimization (PDM, \citealt{1978ApJ...224..953S}) methods for signal search, depending on the characteristics of each light curve. These tools can be classified in two groups, those of the DFT family (DFT, LS, BGLS) based on Fourier transform and those based on phase diagram methods (SL, PDM), which have the advantage of being sensitive even for highly non-sinusoidal signals, as in the case of eclipsing binaries or pulsating stars.

An assessment of the FAP of detecting a signal due to statistical fluctuations of the power spectrum follows \citet[][section III.c]{1982ApJ...263..835S}. The level of Lomb power $z_0$ correspondent to a false alarm probability $p_0$ is 

\begin{equation}
z_0 = -\log \left[ 1 - (1-p_0)^{1/N_f} \right],
\end{equation}

\noindent where $N_f$ is the number of independent frequencies in the Lomb spectrum. For unequally spaced data, $N_f$ is not well defined, but overestimating it increases $z_0$, that is, makes the correspondent power level higher. We surely are overestimating this value, since with $N_f$~=~1~$\times$~10$^6$, the Lomb spectrum is finely sampled.

The uncertainty in period for the detected peaks in the Lomb spectrum, $\sigma_P$, follows the  derivation of \citet{1981kovacs} and is very similar to an expression in \citet{hb1986},

\begin{equation}
\sigma_P = \frac{1}{2 \pi} \sqrt{\frac{24}{N}} \frac{P^2}{T} \frac{\sigma}{R},
\end{equation}

\noindent where $R$ is the semi-amplitude of the signal, $\sigma^2$ is the variance of the noise around the signal, $T$ is the span of the light curve and $N$ is the number of points in it.

The relevant results are presented for each object in Section \ref{sec_classification}.

\section{OBSERVATIONAL CHARACTERISTICS OF MAGNETIC CVS}
\label{sec_classification_criteria}

In this section, we describe the criteria used to classify an object as a mCV candidate. They are broadly 
consistent with previous studies in the field 
\citep[e.g.,][]{2002AJ....123..430S,2003AJ....126.1499S,2004AJ....128.1882S,2005AJ....129.2386S,2006AJ....131..973S,2007AJ....134..185S,2009AJ....137.4011S,2011AJ....142..181S,2014AJ....148...63S,2012AJ....144...81T,2016arXiv160902215T,2014MNRAS.443.3174B,2012MNRAS.421.2414W,2014MNRAS.437..510C}. 

\begin{figure*}
\plotone{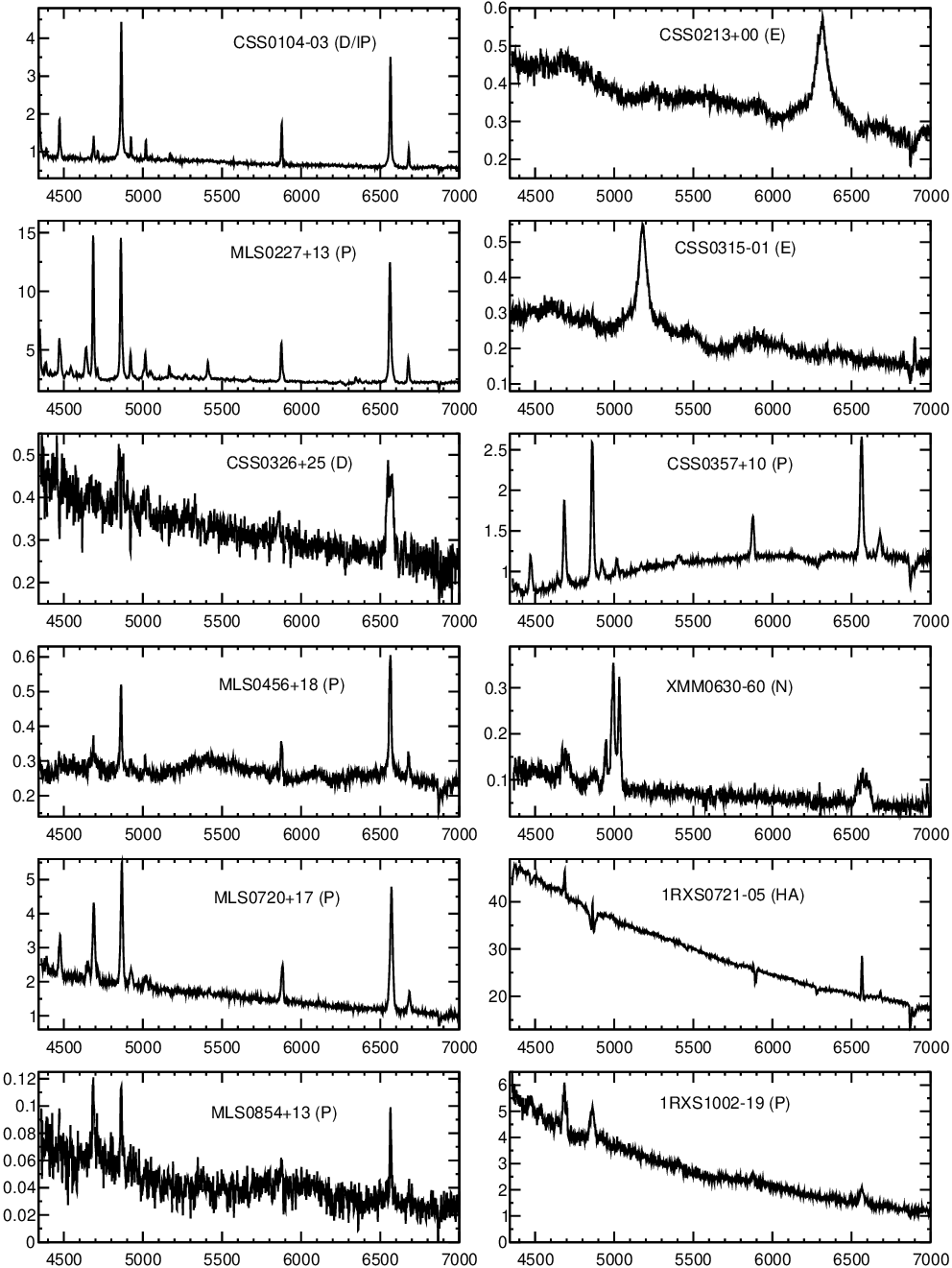}
\caption{Identification spectra arranged by right ascension. The vertical axes are F$_{\lambda}$ in units of $10^{-16}$ erg cm$^{-2}$ s$^{-1}$  \r{A}$^{-1}$ and the horizontal axes are Wavelengths, in \r{A}. \label{allspecs}
}
\end{figure*}

Our spectra, shown in Figure~\ref{allspecs}, allow us to differentiate CVs (Sects.~\ref{sec_polars} to \ref{sec_nova}) from stellar objects without accretion because of the lack of emission lines in the latter. Moreover, active galactic nuclei sources (Sec.~\ref{sec_misc}) are also easily differentiated from CVs because of the redshifted, broad permitted lines. 
On the other hand, the distinction between magnetic and non-magnetic CVs is much more challenging. Hence this section concentrates on the observed characteristics of mCVs.

Usually, the confirmation of magnetic accretion, and therefore the classification as a mCV, cannot be achieved with a single observational technique.
The two kinds of observations that definitively confirm the presence of a magnetic white dwarf are the detection of polarized emission or Zeeman-splitting of spectral lines. But these observations are very time-consuming, particularly for IP systems in which these signals are barely discernible due to the masking caused by the contribution of the disk emission. Hence, to maximize the efficiency of finding mCVs in this work, we performed exploratory spectroscopy of selected objects. The spectral features and other properties of the objects are then analyzed in order to verify if they remain as mCV candidates. The characteristics used to this classification are discussed below. In many cases, a final classification as mCV can only be achieved with complementary observations, which are beyond the scope of this paper.

The mCV spectra typically do not show conspicuous contribution from the stellar components of the system: the continuum is usually flat or blue. When in low state, the optical spectrum of polars may show cyclotron humps \citep[e.g.,][]{2012A&A...546A.104T}, which are not observed in IPs. 

The most conspicuous spectral signature of accretion is the presence of emission lines, mainly of hydrogen Balmer series. However, this is 
a necessary but not sufficient criterion 
for mCVs classification. Magnetic CVs usually present high-ionization lines of \ion{He}{2} and the \ion{C}{3}/\ion{N}{3} blend at 4650 \r{A} originated in the accretion column, although these features may be present in novalike and dwarf nova systems as well \citep{warner1995}.
Specifically, the equivalent width of the hydrogen Balmer and \ion{He}{2} 4686\r{A} lines were proposed as criteria for the classification of magnetic CVs. 
According to \citet{silber1986} -- as cited by Mukai's Intermediate Polars Homepage\footnote{\url{http://asd.gsfc.nasa.gov/Koji.Mukai/iphome/iphome.html}} -- IPs are characterized by H$\beta$ equivalent width larger than 20\AA\ and \ion{He}{2} 4686\r{A}/H$\beta$ larger than 0.4. However, as it will be shown, these quantities should be understood as suggestive of magnetic accretion, but not a {\it sine qua non} condition. 

The line profiles are also useful to the object classification. Polars present narrow spectral lines, due to the absence of an accretion disk. Although low inclination non-magnetic CVs can also display narrow emission lines, polar lines are generally asymmetric, 
because they are composed of distinct components originated from different parts of the accretion flow and from the illuminated secondary face. IPs, however, can present line profiles very similar to those of non-magnetic CVs, with large widths and double peaks. 

We have searched for high-energy counterparts because magnetic CVs are frequently 
detected in X-ray surveys.
In fact, a large number of magnetic CVs have been discovered by X-ray all-sky surveys, as ROSAT \citep[e.g.,][]{1993AdSpR..13..115B}. 
Magnetic CVs have higher X-ray to optical flux ratios than dwarf novae in quiescence \citep[e.g.,][]{1993AdSpR..13..115B,2011PASP..123.1054F,2011AJ....142..181S}. This information and the hardness ratios, for objects having ROSAT counterparts, were used to provide a hint on whether an object might be a mCV \citep[e.g.,][]{2014OEJV..167....1H}. However, some confirmed mCVs are not bright enough to be detected by the ROSAT survey \citep[e.g.,][]{2015MNRAS.451.4183S,2014ASPC..490..389M}. Figure 2 of \citet{1993AdSpR..13..115B} shows that polars with the lowest X-ray to optical flux ratio can be missed by ROSAT if they have V $>$ 19 mag. The limit for IPs is even more restrictive: V $>$ 17~mag.

From a photometric point of view, polars alternate high and low brightness states in optical and X-rays with drops of $1-3$ magnitudes, probably associated to variations in the mass-transfer rate. They do not present outbursts, unlike some IPs. In short time-scales, mCV's fluxes vary periodically with the WD spin. The modulation of polars can have an amplitude as high as 1-2 magnitudes and their periods are around hours because the WD spin is synchronized with the orbital revolution. IPs have smaller amplitudes (tenths of magnitudes) and smaller periods (minutes). We searched for periodic variability in all stellar objects, as described in Sec.~\ref{sec_time_analysis}. However, the shape of the light curves vary from cycle to cycle due to flickering, which can hamper the discovery of periodic variability in long-term and sparse photometric data, as CRTS.

The summary of the above discussion is that polars are more easily identified in our sample than IPs, because polars lack an accretion disk in such way that the accretion column dominates their emission. In contrast, IPs have a large contribution from the disk, which makes them observationally similar to non-magnetic CVs. Moreover, the X-ray emission and high-ionization emission lines are not exclusive properties of mCVs while, on the other hand, some mCVs are not detected in X-rays \citep[e.g.,][]{2015MNRAS.451.4183S,2014ASPC..490..389M}. Therefore, these observables were considered in the proposed classification of each object, but they are not definite criteria.

\section{CLASSIFICATION RESULTS}
\label{sec_classification}

\begin{figure*}
\plotone{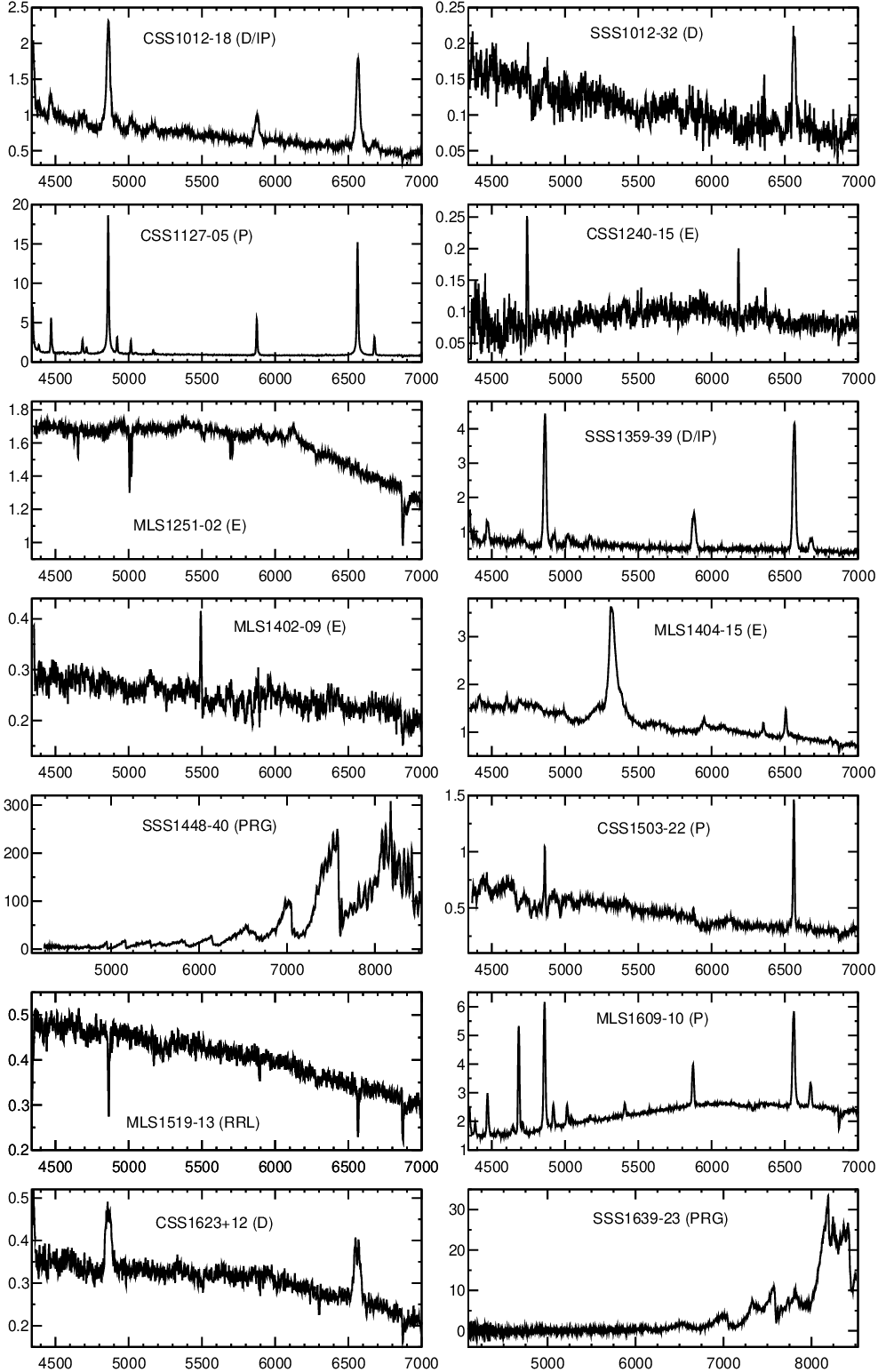}
\flushleft
    { Fig. 1.---\it{Continued}} 
\end{figure*}

\begin{figure*}
\plotone{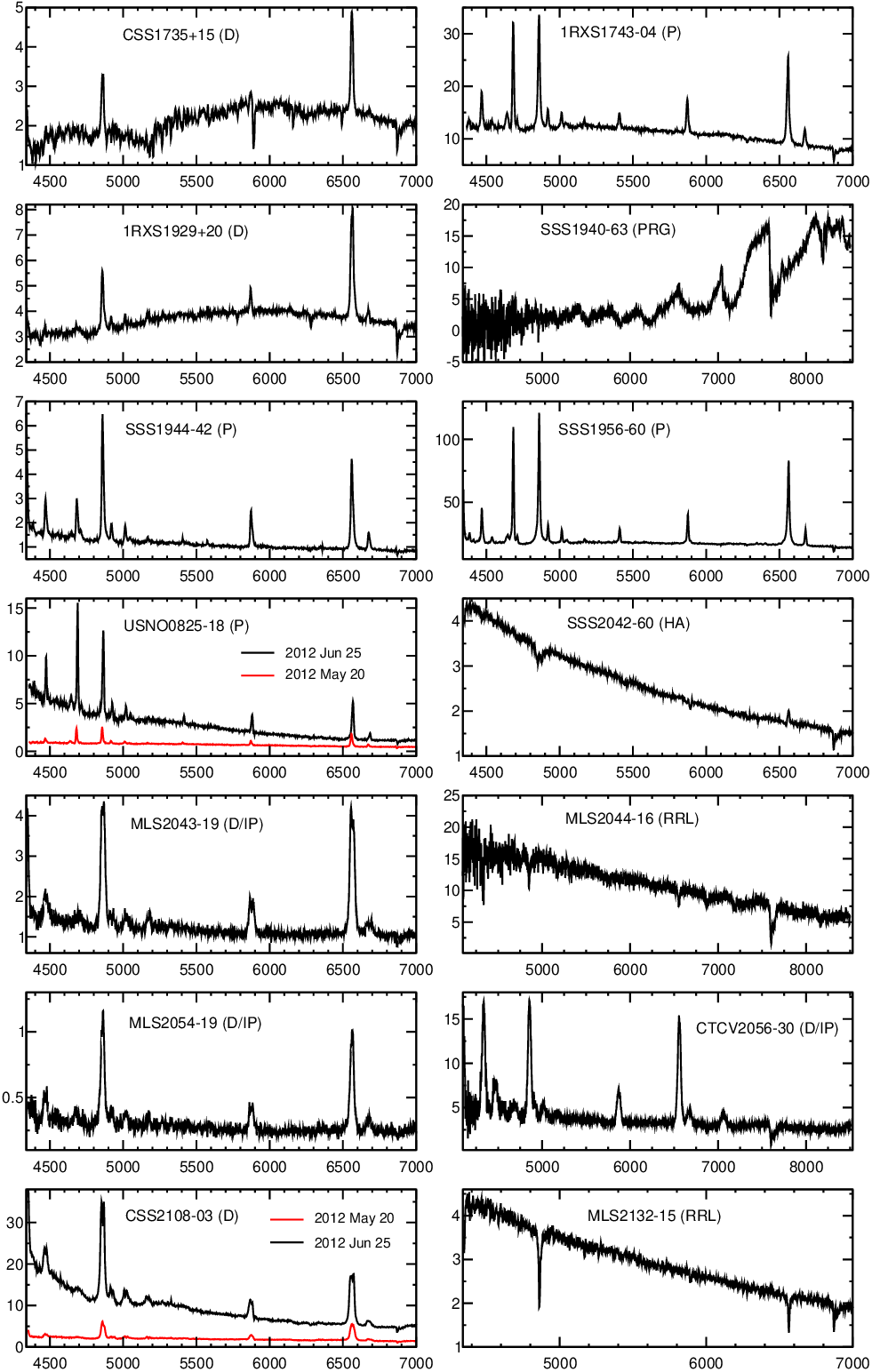}
\flushleft
    { Fig. 1.---\it{Continued}}
\end{figure*}

\begin{figure*}
\plotone{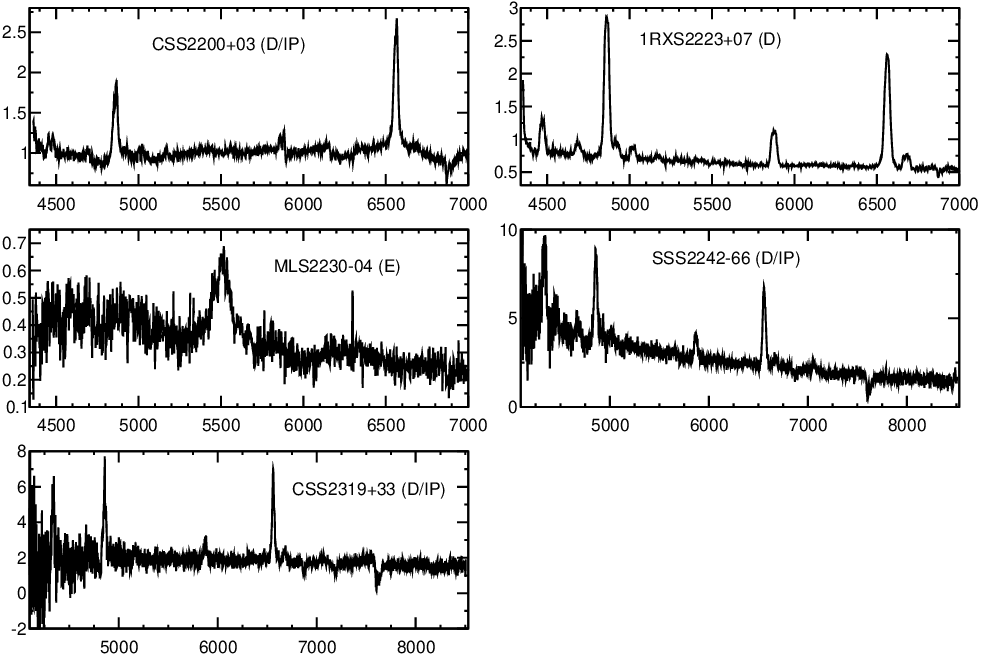}
\flushleft
    { Fig. 1.---\it{Continued}}
\end{figure*}

\subsection{Polar Candidates}
\label{sec_polars}

This section comprises the polar candidates, or diskless accretors. The other mCVs - IP candidates - are placed in the next section together with other disk-accretion systems. Table \ref{fwhm} presents the FWHM and the equivalent width of the Balmer and \ion{He}{2} 4686~\r{A} lines of the objects classified as CVs in this work.

\subsubsection{MLS0227+13 (MLS110213:022733+130617)}

Its CRTS light curve ranges from 16.5 to 19.5 mag on timescales of days. The spectrum is typical of polars with H$\beta$ more intense than H$\alpha$. The \ion{He}{2} 4686~\r{A} emission line is as intense as H$\beta$ (Table \ref{fwhm}) and the high-ionization Bowen \ion{C}{3}/\ion{N}{3} is also present. In our follow-up observational study, \citet{2015MNRAS.451.4183S} confirmed MLS0227+13 as an eclipsing polar with an orbital period of 3.787~h, using polarimetric, photometric and spectroscopic data. The system has no X-ray counterpart, which can be explained by its large distance from us \citep{2015MNRAS.451.4183S}.

\subsubsection{CSS0357+10 (CSS091109:035759+102943)} 

CSS0357+10 was detected as a transient by CRTS on 2009 November 9. The V magnitudes are highly variable, spanning from 16.5 to 21 mag in short timescales. \citet{2012AN....333..717S} performed a time-resolved photometric study of this target and found a periodicity of 114 min with 0.8 mag amplitude, interpreted as the orbital period of a non-eclipsing polar candidate. Recently, \citet{2016arXiv160902215T} obtained spectroscopic time series of this source and corroborated the 114 min orbital period. This object is also detected in X-rays (ROSAT, XMM-Newton), presenting large variability \citep{2012AN....333..717S}.

The identification spectrum is dominated by intense Balmer and \ion{He}{2} 4686 \r{A} lines. It also presents \ion{He}{1} and  weak \ion{He}{2} 5411 \r{A}, Bowen \ion{C}{3}/\ion{N}{3} complex and \ion{Fe}{2} 5170 \r{A} emissions. The Balmer and \ion{He}{2} 4686 \r{A} lines are asymmetric, with extended red wings, and H$\beta$ is more intense than H$\alpha$ (Table \ref{fwhm}), confirming the polar classification.

\subsubsection{MLS0456+18 (MLS101203:045625+182634)}

This object was detected as a variable star on 2013 December 10 by the CRTS. Its light curve spans from 18.5 to 21.5 mag, reaching 1 mag variation amplitude in timescales of hours. Narrow Balmer emission lines dominate the identification spectrum, which also presents \ion{He}{1} and \ion{He}{2} 4686 \r{A} lines.
The latter line consists of a broad component (EW$ = 18$ \r{A}) superimposed on a very narrow central emission (EW$ = 1$ \r{A}). The composite line profile, the EW of H$\beta$, and the ratio between \ion{He}{2} 4686 \r{A} and  H$\beta$ (Table \ref{fwhm}) suggest a mCV classification. The continuum shows humps that may be associated to cyclotron harmonics. This property and the absence of an X-ray counterpart are consistent with a low accretion-rate polar or a pre-polar system \citep[e.g.,][]{2005ApJ...630.1037S}. The power spectrum of the CRTS data has its highest peak at 0.09851004~d (2.36~h), which has a FAP smaller than 0.0001 (=0.01\%). The formal error of this period is $1.3~\times~10^{-7}$~d (see Section \ref{sec_time_analysis}). The corresponding phase diagram (Figure 2) is compatible with the light curve of a polar system. This period places MLS0456+18 inside the period gap of CVs.  
As several objects in this work, a definite classification demands more observations.

\begin{figure}
\plotone{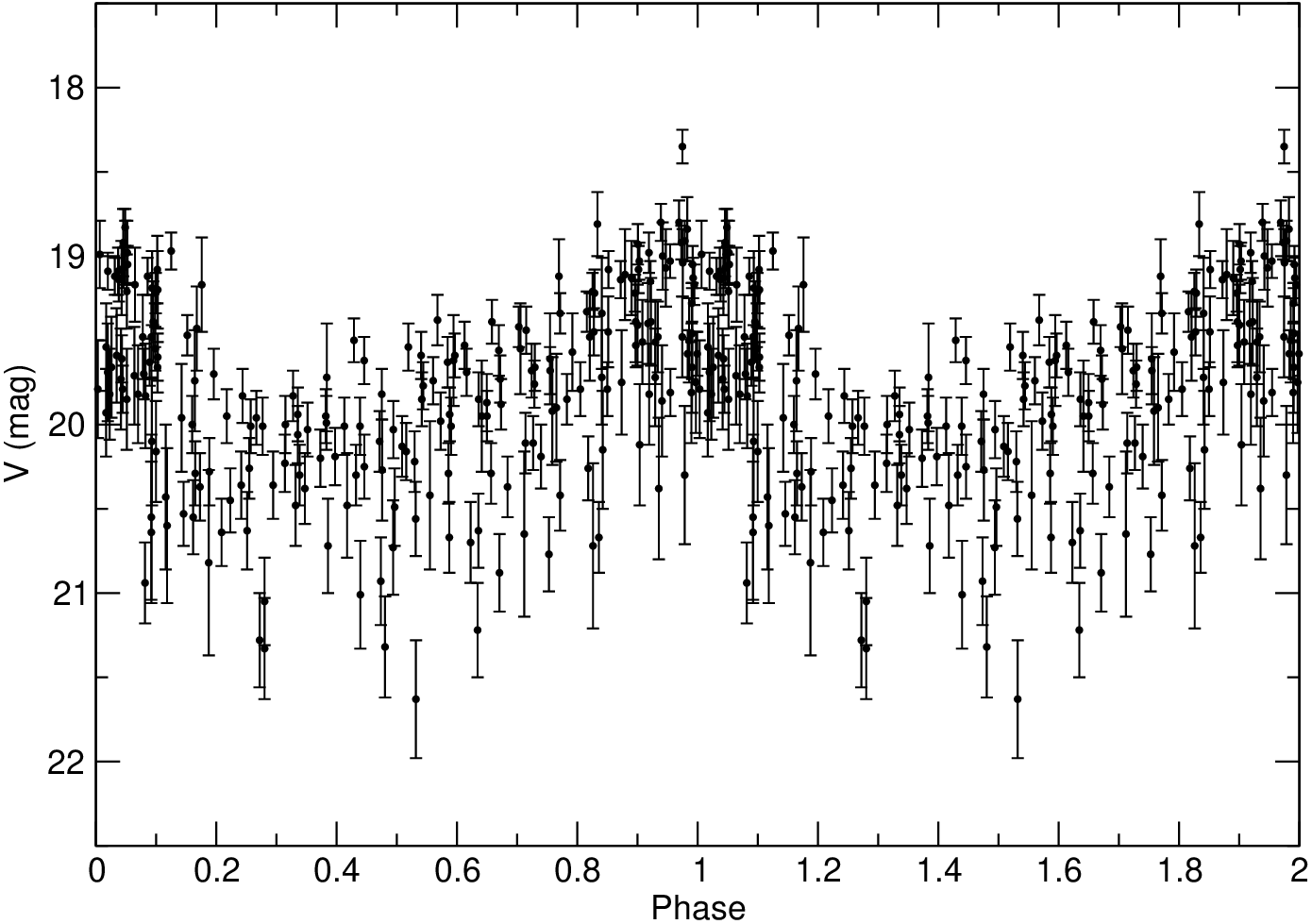}
\caption{CRTS light curve of MLS0456+18 folded with the period of 0.09851004 d. } \label{figmls0456}
\end{figure}

\subsubsection{MLS0720+17 (MLS101226:072033+172437)}

The CRTS light curve of MLS0720+17 presents 0.5 mag amplitude variability around 18.5 mag, occasionally dropping to 20-21 mag, possible during eclipse events \citep{2014MNRAS.441.1186D}. The Goodman identification spectrum is typical of polars, with Balmer, \hheii, \ion{C}{3}/\ion{N}{3} and \ion{He}{1} narrow emission lines superimposed on a continuum with a blue slope. The \ion{He}{1} line profiles are asymmetric with extended blue wings in this spectrum. No X-ray source was found near its coordinates. The time series analysis applied to the CRTS data does not show any periodic signal that could explain the excursions to lower fluxes. \citet{2016arXiv161107885H} have recently presented a light curve spanning approximately 1h. It shows a modulation that the authors interpret as an eclipse. We consider it can arise from cyclotron emission, as usually seen in polars. Moreover, the narrow spectral lines are inconsistent with an eclipsing system having a disk.

\subsubsection{MLS0854+13 (MLS120127:085402+133633)}

This target matches a blue stellar counterpart in the SDSS, with 
$u=20.18$, $g=20.58$, $r=20.43$, $i=20.28$, and $z=20.06$ in DR8. It presents two brightness levels with averages 19.7 and 21.5~mag in the CRTS data. It is associated with the X-ray source 1WGA J0854.0+1336 in the WGA Catalog of ROSAT Point Sources\footnote{\url{http://heasarc.gsfc.nasa.gov/wgacat/}}. This object is also classified as BOSS J085402.10+133632.8, a z=2 QSO in the MILLIQUAS Catalog V4.7\footnote{\url{http://heasarc.gsfc.nasa.gov/w3browse/all/milliquas.html}}, with no SDSS spectrum available. However, our spectrum is completely consistent with a Galactic object, being similar to the spectrum of a polar in low accretion state, with cyclotron humps and emission lines of \hheii, H$\beta$ and H$\alpha$ on a blue continuum. No periodic signal was found in the CRTS time series analysis.

\subsubsection{1RXS1002-19 (1RXS J100211.4-192534)}

This source was discovered by ROSAT \citep{1993AdSpR..13..115B} and classified as a polar with an X-ray orbital period of 106 min \citep{1994ASPC...56..119B}. 1RXS1002-19 was observed with XMM-Newton \citep{2003MNRAS.338..219R} and the resulting X-ray light curve suggests that the source of radiation is located in one hemisphere of the WD, while modeling of the X-ray spectrum yields a WD mass close to 0.5 M$_\odot$. 

Our optical spectrum of 1RXS1002-19, the first for this source, has a strong blue continuum. The \heii line is more intense than H$\beta$, which in turn is more intense than H$\alpha$ (Table \ref{fwhm}). \ion{He}{1} features are also present in emission. The FWHM of the H$\beta$ and \heii lines are around 2000~km s$^{-1}$ (Table \ref{fwhm}), which is unusual for a polar. The equivalent width of the H$\beta$ emission is 11~\r{A}, smaller than the value of 20~\r{A} from the Silber's criterion \citep{silber1986} for a magnetic CV classification.
Our spectrum converts to V~=~18~mag. This locates the object in the polar systems region of X-ray count rate versus optical mag diagram, according to Figure 2 of \citet{1993AdSpR..13..115B}.

\subsubsection{CSS1127-05 (CSS110225:112749-054234 = SSS120624:112749-054236)}

CSS1127-05 has a blue point source counterpart in the SDSS, with $u=20.00, g=20.12, r=20.09, i=20.28$ and $z=20.15$ in DR8. The CRTS light curve clearly shows two distinct brightness states with average magnitudes of 18.5 and 20.5. The exploratory spectrum has high signal to noise ratio, displaying strong and narrow Balmer lines in emission on a flat continuum, besides less intense \heii and \ion{He}{1}. H$\beta$ is more intense than H$\alpha$. The H$\alpha$ EW (204~\AA) is the largest among the objects we classify as CV (Table~\ref{fwhm}). In spite of the low ratio between \heii and H$\beta$ (0.16), considering the Silber's criterion \citep{silber1986}, we suggest it is a polar due to the presence of high and low-brightness states. However, we do not discard a possible IP classification since these objects may also present low states, although less commonly. No significant periodicity was found in the CRTS data. The system has no X-ray counterpart.

\subsubsection{CSS1503-22 (CSS100216:150354-220711 = SSS100512:150354-220710)}

The light curve of CSS1503-22 presents two alternating brightness states at average magnitudes 19.5 and 17.5, both states with dispersion of 1 to 1.5 mag. \citet{2012MNRAS.421.2414W} performed photometric monitoring of CSS1503-22 in March and April 2010 on five runs summing up 15~h of observations when it was at high state. They found a period of 133.38 min and, from its photometric behavior and the detection as a Swift X-ray source, they suggested it to be a polar. 
The flux of our SOAR spectrum corresponds to 19.7~mag, which is consistent with the CRTS low state. 
The spectrum presents absorption features from 4500 to 6000~\r{A}, which we interpret as from the stellar components. Some absorptions occur at typical wavelengths of TiO bands from a M2 -- M5 secondary star (see, for instance, \citealt{2002A&A...383..933M} and \citealt{2009ssc..book.....G}). The absorptions near H$\beta$ and to the blue of H$\alpha$ are probably due to Zeeman splitting in the photosphere of the white dwarf. See, for instance, the SDSS~J214930.74-072812.0 spectrum and the superimposed model atmosphere with B = 44.7~MG presented in \citet{2009A&A...506.1341K}. The spectrum also displays narrow Balmer lines, possibly with blue wings, and H$\beta$ being half the intensity of H$\alpha$ (Table \ref{fwhm}). The \ion{He}{1} 5876 \r{A} line is also present. The Zeeman absorptions hamper the detection of possible \heii or Bowen complex emissions. The EW of H$\beta$ is 16~\r{A}, slightly smaller than the value of 20~\r{A} from the Silber's criteria \citep{silber1986}, but
the accuracy of the measured EW can be affected by the cited absorptions.

\subsubsection{MLS1609-10 (MLS110609:160907-104013)}

\begin{figure}
\plotone{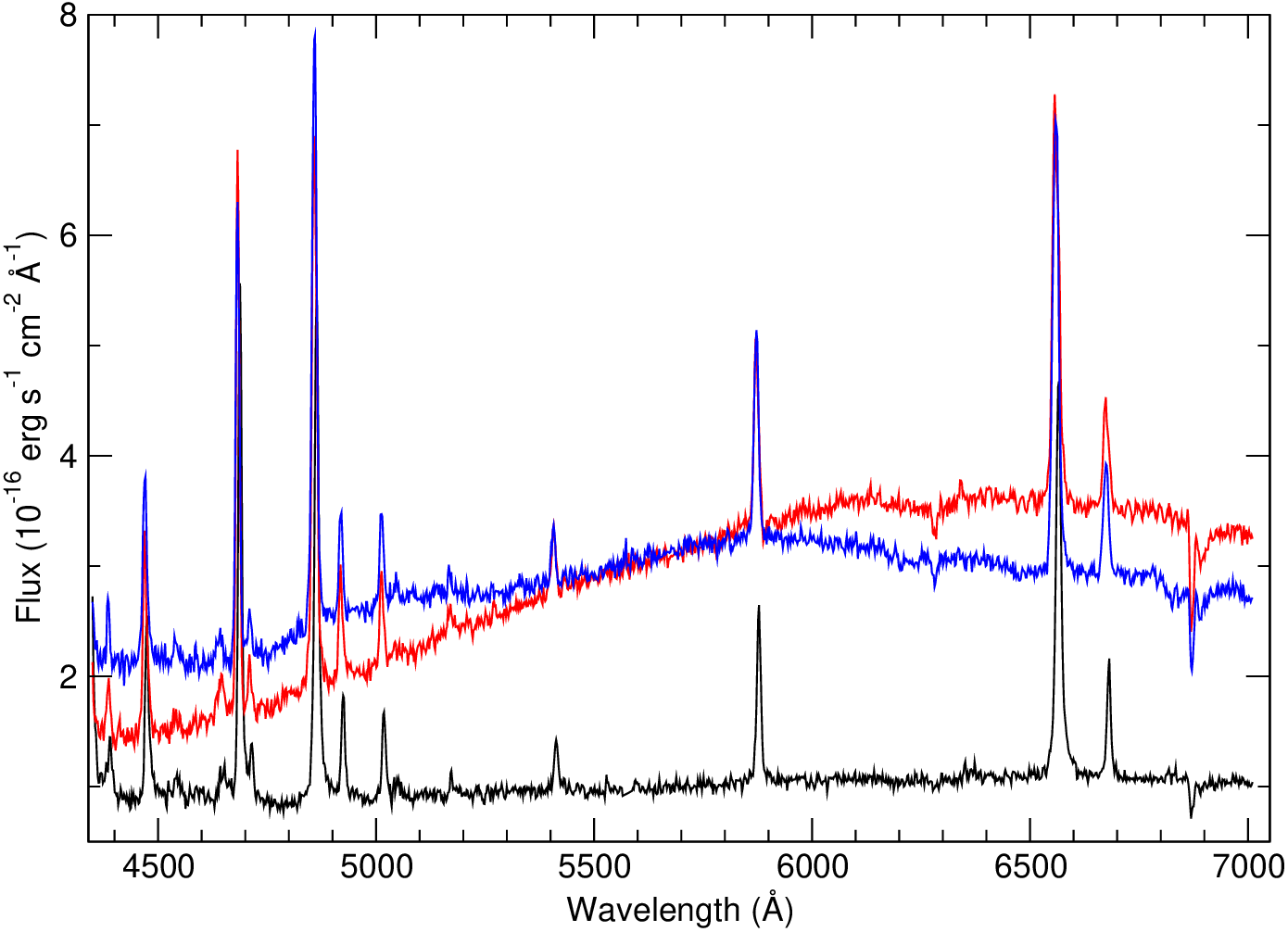}
\caption{Three consecutive (black, red, blue) spectra of MLS1690-10 obtained with 20 min intervals.} \label{mls1609specs}
\end{figure}

The CRTS light curve of this object has measurements scattered from 16.8 to 20.5~mag. The exploratory spectrum presents 
intense and narrow Balmer lines in emission. These lines are slightly asymmetrical, presenting extended red wings. H$\beta$ is more intense than H$\alpha$, while \heii is almost as intense as H$\beta$ (Table \ref{fwhm}). The Bowen complex at 4630-4650 \r{A} is clearly visible. 
The spectrum shown in Figure \ref{allspecs} is the average of three consecutive spectra, obtained with 20 min intervals. They show a remarkable variability in the continuum, hence they are individually presented in Figure \ref{mls1609specs}.
We performed a 3.6 h span photometric time-series of this source on 2014 April 28, using the 1.6 m P\&E telescope at OPD and the Andor iKon-L936BV camera, obtaining 39 images with 240 s exposure time through the V filter. The differential photometry was converted to V magnitude using as reference the star NOMAD1~0793-0295201, with $m_v=16.51$ \citep{2004AAS...205.4815Z}. 

\begin{figure}
\plotone{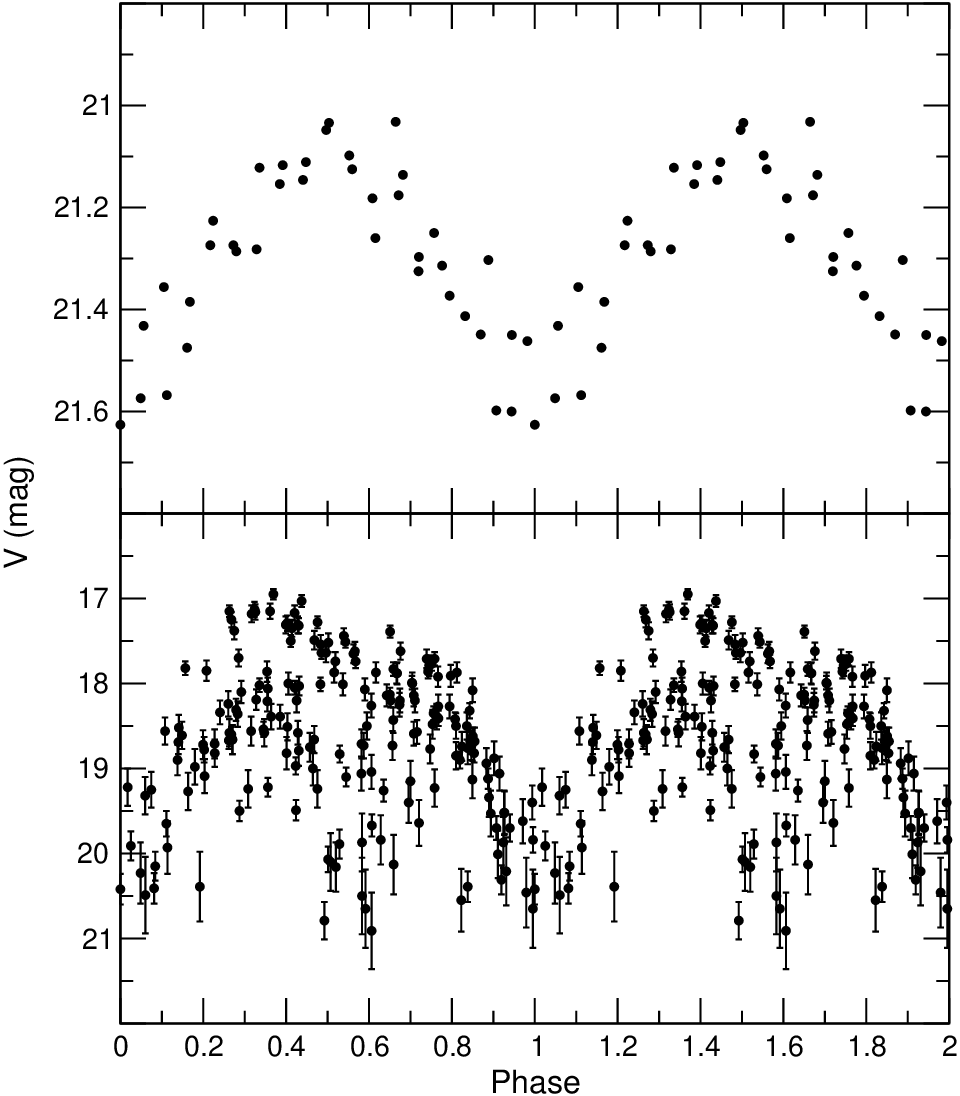}
\caption{OPD (upper panel) and CRTS (lower panel) light curves of MLS1609-10, both folded with the period of 0.075439 d.} \label{figmls1609-10}
\end{figure}

The OPD lightcurve is consistent with a polar classification and shows a modulation with a period around 0.077~d and 0.60$\pm$0.05 mag amplitude. In the search for periodic signals in the CRTS data, the PDM method yielded a period of P~=~0.075439 $\pm$ 0.000001~d, compatible with the OPD period. Figure \ref{figmls1609-10} shows both OPD and CRTS light curves folded with P~=~0.075439~d and epoch T$_{min}=2~456~776.7108$ HJD. The average magnitude of the OPD data is V=21.3, therefore the source was caught in a very low state, whereas the individual spectra correspond to 17.8 and 19 V mag. The system has no X-ray counterpart.

\subsubsection{1RXS1743-04 (1RXS J174320.1-042953 = USNO-B1.0 0855-0326594)}

This CV was discovered by \citet{2011AstL...37...91D} among X-ray sources in the ROSAT catalog. \citet{2012PZ.....32....3D} performed time-resolved photometry and suggested the target to be a polar or IP, based on the amplitude (0.8 mag in white light) and shape of the light curve.
They determined an orbital period of 2.08 h. 

Our exploratory spectrum is typical of polars, with very intense \heii and 
H$\beta$ more intense than H$\alpha$. The \ion{C}{3}/\ion{N}{3} complex is also present and the continuum increases to the blue. Our spectrum corresponds to V~=~16.3~mag.

\newpage

\begin{deluxetable}{cccccccc}
\tabletypesize{\scriptsize}
\tablecaption{Full Width at Half Maximum (FWHM\tablenotemark{a}) and Equivalent Width (EW) of \hheii, H$\beta$ and H$\alpha$ emission lines of the objects classified as polars and disk systems. \label{fwhm}}
\tablewidth{0pt}
\tablehead{
\colhead{Object} & \colhead{FWHM He II} & \colhead{$-$EW He II } & \colhead{FWHM H$\beta$ } & \colhead{$-$EW H$\beta$ } & \colhead{FWHM H$\alpha$} & \colhead{$-$EW H$\alpha$} & \colhead{He II/H$\beta$}\\
\colhead{} & \colhead{(km s$^{-1}$)} & \colhead{({\AA}) } & \colhead{(km s$^{-1}$) } & \colhead{({\AA})} & \colhead{(km s$^{-1}$)} & \colhead{({\AA}) } & \colhead{}
}
\startdata
CSS0104-03  & 490  & 6   & 580  & 59 & 450   & 63 & 0.32     \\
MLS0227+13  & 630  & 47  & 700  & 66 & 600   & 76 & 1.01      \\
CSS0326+25  &  \nodata & \nodata & p\tablenotemark{b}  & p  & 2380:  & 47: & \nodata        \\
CSS0357+10  & 1040 & 22  & 1050 & 40 & 900   & 30 & 0.73        \\
MLS0456+18  & 750  & 13  & 730  & 20 & 690   & 23 & 0.71        \\
MLS0720+17  & 1040 & 19  & 940  & 32 & 850   & 65 & 0.79  \\
1RXS0721-05 & 530  & 1   & 420  & 1  & 400   & 4  & 0.83  \\   
MLS0854+13  & 930  & 15  & 810  & 24 & 490   & 30 & 1.06   \\
1RXS1002-19 & 1850 & 12  & 2090 & 11 & 1200  & 10 & 1.17     \\
CSS1012-18  &   p  & p   & 1710 & 64 & 1250  & 78 & \nodata   \\
SSS1012-32  & \nodata  & \nodata & p & p & 1120:  & 53: & \nodata   \\
CSS1127-05  & 540  & 14  & 550  & 183& 450   & 204& 0.16   \\
SSS1359-39  & p    & p   & 1240 & 157& 1030  & 197& \nodata  \\
CSS1503-22  & \nodata  & \nodata & 490  & 16 & 490   & 44 & \nodata  \\
MLS1609-10  & 730  & 29  & 870  & 40 & 840   & 26 & 0.87   \\
CSS1623+12  & \nodata  & \nodata & 2670 & 19 & 2370  & 28 & \nodata   \\
CSS1735+15  & \nodata  & \nodata & 1250 & 23 & 840   & 24 & \nodata   \\
1RXS1743-04 & 760  & 21  & 880  & 32 & 730   & 36 & 0.96 \\
1RXS1929+20 & p    & p   & 1000 & 14 & 1020  & 30 & \nodata    \\
SSS1944-42  & 950  & 16  & 890  & 65 & 770   & 65 & 0.46   \\
SSS1956-60  & 600  & 48  & 700  & 80 & 570   & 59 & 0.91  \\
USNO0825-18 (May 20) & 630  & 22  & 660  & 22 & 630   & 42 & 0.97  \\
USNO0825-18 (Jun 25) & 580  & 27  & 690  & 26 & 640   & 38 & 1.23   \\
SSS2042-60  & \nodata  & \nodata & 860  & 1  & 630   & 2  & \nodata \\
MLS2043-19  & p    & p   & 2070 & 87 & 1650  & 91 & \nodata\\
MLS2054-19  & p    & p   & 1700 & 84 & 1311  & 93 & \nodata  \\
CTCV2056-30 & p    & p   & 2430 & 124& 1850  & 160& \nodata \\ 
CSS2108-03 (May 20) & p    & p   & 1720 & 51 & 1520  & 81 & \nodata \\
CSS2108-03 (Jun 25) & p     & p   & 1930 & 66 & 1690  & 92 & \nodata \\
CSS2200+03  & p    & p   & 2010 & 32 & 1320  & 43 & \nodata \\
1RXS2223+07 & p    & p   & 2010 & 90 & 1720  & 128& \nodata  \\
SSS2242-66 (Sep 03) & p  & p   & 2200 & 71 & 1890  & 125& \nodata   \\
SSS2242-66 (Sep 04) & p  & p   & 2300 & 56 & 1610  & 62 & \nodata  \\
CSS2319+33  & p:  & p: & 1360 & 144& 1130  & 85 & \nodata     \\
\enddata
\tablenotetext{a}{The FWHM were measured from Voigt profile fiting. }
\tablenotetext{b}{The symbol p means that the line is present in the spectrum but its S/N is low and no measurements were possible. }
\end{deluxetable}

\subsubsection{SSS1944-42 (SSS100805:194428-420209)}

The CRTS light curve of this object presents two brightness levels separated by 2 mag and short timescale variability. \citet{2014MNRAS.437..510C} performed photometric monitoring during the bright state and classified it as a polar from its light curve, which is modulated with a period of 1.53~h. Our spectrum was obtained in the low-brightness state (V $\approx$ 19~mag) and presents narrow, single peaked Balmer, \ion{He}{1} and \heii emission features over a blue continuum. The Balmer and \ion{He}{2} lines have extended red wings and H$\beta$ is 35\% more intense than H$\alpha$ (Table \ref{fwhm}). The \ion{C}{3}/\ion{N}{3} Bowen complex at 4640 \r{A} and the lines of \ion{Fe}{2} 5169 \r{A} and \ion{He}{2} 5412 \r{A} are also visible. The system has no X-ray counterpart.

\subsubsection{SSS1956-60 (SSS110526:195648-603430)}

This source presents many kinds of variability in the CRTS light curve: long (years) and short (one month) term variations between magnitudes 18 and 15, and 1 mag variations in timescales of a few days. SSS1956-60 is classified as the novalike system CV V348~Pav \citep{2001PASP..113..764D} and is also identified as V1956-6034, a polar candidate by its spectrum \citep{1994AJ....107.2172D}.  Our exploratory spectrum displays narrow Balmer, \ion{He}{1} and \ion{He}{2} emission lines. The line of \heii is almost as intense as H$\beta$, which in turn is more intense than H$\alpha$ (Table~\ref{fwhm}). Also present in the spectrum are the lines of the Bowen complex, \ion{He}{2} 4541 and 5412 \r{A} and \ion{Fe}{2} 5169 \r{A}. In the spectrum obtained in 1992 by \citet{1994AJ....107.2172D}, the ratio of the intensities H$\beta$/\heii is 0.25, much smaller than the value of 0.91 we measured in our spectrum from 2012. An X-ray counterpart was detected with 1.8 counts sec$^{-1}$ total band count rate at these coordinates by XMM-Newton in the slew data mode, and was registered as XMMSL1 J195647.7-603425 in the XMM-Newton Slew Survey Clean Source Catalog \citep{2008A&A...480..611S}.

\subsubsection{USNO0825-18 (USNO-A2.0 0825-18396733)}

The discovery of this polar candidate is reported by \citet{2010IBVS.5952....1K}. From the R light curve, they obtained an orbital period of 0.0848~d (2.035~h), and their spectrum shows \heii almost as  intense as H$\beta$ and also \ion{He}{1} and the Bowen blend. 
\citet{2015AstBu..70..328A} performed polarimetric observations of this source and found variable circular polarization reaching $-31\%$.

We obtained exploratory spectra of USNO0825-18 in two occasions in 2012, 35 days apart. In the second run, the flux in the blue was 6 times more intense than in the first observation (Figure \ref{allspecs}) and the \heii to H$\beta$ intensity ratio increased from 0.97 to 1.23 (Table~\ref{fwhm}).

\subsection{Systems with Disks}

In this section, we present the systems with accretion disks. Possible IPs are also included here.

\subsubsection{CSS0104-03 (CSS091009:010412-031341)}

This object was discovered and classified as an eclipsing CV by CRTS on 2009 October 9 in a single outburst when it reached V=17.2 mag. CRTS data span from 2005 to 2015, during which the object showed average V=19 mag and $\sim1$ mag variability in timescales of hours, with occasional 21 mag excursions. Another outburst was recorded by ASAS at 16.39 mag on 2013 August 27 \citep{vsnet16294}.
\citet{2014AJ....148...63S} present a spectrum with strong emission Balmer lines obtained on 2013 September 3 and classify the object as a dwarf nova, possibly eclipsing due to the dispersion of the CRTS quiescent light curve. This spectrum was taken in a state brighter than the quiescent flux in the CRTS light curve. The SDSS DR9 counterpart of CSS0104-03 presents $u=19.2, g=19.5, r=19.1, i=18.8$ and $z=18.9$.    

Our Goodman spectrum shows narrow and single peaked Balmer, \ion{He}{1} and \heii emission lines, H$\beta$ being 1.26 times more intense than H$\alpha$ (see Table \ref{fwhm}). The continuum is flat with a slight slope to the blue. The Balmer lines have extended wings and are slightly asymmetrical. The average continuum flux in our spectrum is a factor of 4 less intense than the continuum flux in the spectrum published by \citet{2014AJ....148...63S}, in which the line of \ion{He}{2} 4686~\r{A} is only marginally detected. CSS0104-03 has the highest \ion{He}{2} 4686~\r{A} EW among our sample of disk-accretion systems. Our spectrum has a flux level around 19 mag in the V band, hence it is consistent with the mean level of the CRTS light curve. 

Our time series analysis of the CRTS data does not reveal any period which could explain the low flux excursions as eclipses. The small widths of the lines are also inconsistent with an eclipsing system. So the system could be a near face-on dwarf nova. However, this classification does not explain the low flux measurements, the two-component emission lines, and the \heii line behavior, since its intensity was expected to be higher at the spectrum from \citet{2014AJ....148...63S} taken at a brighter state. The object has no X-ray counterpart. If we suppose that its low optical brightness is due to its distance, it is possible that its X-ray luminosity is not high enough to allow the object detection in X-ray surveys. This seems to be true for MLS0227+13 \citep{2015MNRAS.451.4183S}. Therefore, a magnetic classification is a reasonable possibility taking into account the observations. Moreover, if the brightest points correspond to eruptions, it could be an IP system, possibly similar to GK~Per or EX~Hya. Spectroscopic time series would be useful to a definite classification of this system.

\subsubsection{CSS0326+25 (CSS090922:032603+252534)}

The light curve of this source, obtained by CRTS, has a large amplitude of variation, from 18.5 to 20 mag. It might be originated by flips between two brightness states. The average spectrum presents broad emission lines of H$\alpha$ and H$\beta$. H$\alpha$ shows a profile with two peaks, while H$\beta$ is quite peculiar, with a strong central absorption at the rest position of the line, in the center of the broad emission.  Also marginally detectable are absorption lines of the \ion{Na}{1} doublet at 5890 -- 5896 \r{A} and the diffuse interstellar band (DIB) at 6284 \r{A}. Our time series analysis does not reveal any remarkable signal. This object has no evidence of being a magnetic system.

\subsubsection{CSS1012-18 (CSS120324:101217-182411 = SSS110203:101217-182411)}

This CRTS source has been discovered by SSS on 2011 February 3, receiving the ID  SSS110203:101217-182411 and was later detected by CSS, on 2012 March 24, when it was assigned as CSS120324:101217-182411. The CRTS light curves varies between 18-20 mag with sporadic outbursts reaching 16-17 mag. Our spectrum displays a blue continuum and broad \ion{He}{1} and Balmer emission lines,  H$\beta$ being more intense than H$\alpha$. \heii is present as well. The spectrum flux corresponds to approximately 19.3~mag in the V band, consistent with the quiescent flux level.
\citet{2016MNRAS.456.4441C} classify it as a dwarf nova with 3 detected outbursts.
Its spectrum resembles the spectrum of the IP candidate CTCV2056-30 (see section~\ref{ctcv}) and short-period IPs like SDSS J2333 \citep{2007MNRAS.378..635S,2005AJ....129.2386S} and HT Cam \citep{2002PASP..114..623K}, so we suggest that it is a member of this class. The String-Length spectrum of the CRTS data shows some structure, but they do not translate into discernible features in the folded diagrams.

\subsubsection{SSS1012-32 (SSS110504:101240-325831)}

This object is associated with the X-ray source 1RXS J101239.6-325839. The CRTS light curve has short term (days) variability between 17.3 and 19.5 mag, but no periodicity was found. 
Moreover, the CRTS light curve does not present DN-type eruptions. The spectrum has low S/N, but H$\alpha$ is clearly seen as a broad emission line while H$\beta$ is weak. The continuum is quite blue. Despite not showing spectral features characteristic of magnetic systems, its ROSAT colors match those sources.
The spectrum shows modulations that might be due to absorptions features of a red stellar component, but we could not associate their wavelengths to features usually found in CV secondaries. We do not classify this system as a mCV because of the lack of spectral evidence.

\subsubsection{SSS1359-39 (SSS110724:135915-391452)}

SSS1359-39 was discovered by CRTS and its light curve has a quiescent brightness level at 19 mag with a dispersion of 1 mag, and sporadic outbursts at magnitudes 14 - 15. The periodicities search had negative results. This source is associated to the X-ray source 1RXS~J135915.6-391447. The exploratory spectrum 
shows features related to an accretion disk, with intense and broad Balmer emission lines (FWHM $>$ 1000 km s$^{-1}$ -- Table \ref{fwhm}) along with \ion{He}{1}, \ion{Fe}{1}, \ion{Fe}{2} and \hheii, and is very similar to the spectrum of the IP candidate CTCV2056-30 (section~\ref{ctcv}), so we classify it as a possible short-period IP.

\subsubsection{CSS1623+12 (CSS080606:162322+121334)}

Photometric monitoring from CRTS yields a flat light curve at 17 mag with rapid and short occasional decreases of 3 magnitudes. SDSS registers this object as a blue point source with $u = 21.85, g = 21.88, r = 21.42, i = 21.38$ and $z = 21.25$. \citet{2014MNRAS.441.1186D} obtained a spectrum of this object in high-accretion state, with broad Balmer absorptions on a blue continuum. Our SOAR spectrum was obtained in a faint state (V $\approx$ 20.2~mag) and shows the Balmer lines in emission. These lines are broad with double peak profiles, indicating a possible high orbital inclination. The system may be eclipsing, which would explain the rapid and deep drops in the CRTS light curve. The time series analysis of the CRTS data does not reveal any clear periodicity, probably because most of the fainter points are upper limits and are not included in the photometric CRTS data. If the object is indeed eclipsing, our spectrum (exposure time of 2700~s) would have been taken predominantly during the eclipse, which should be relatively long. This source could be alternatively interpreted as a VY~Scl novalike variable after its spectral and photometric behavior.

\subsubsection{CSS1735+15 (CSS110623:173517+154708)}

The light curve obtained by CRTS is nearly flat around 17 and 17.5 mag with one episodic measurement at 14.3 mag and another at 18 mag. \citet{2012AJ....144...81T} obtained one spectrum in September 2011 in which H$\alpha$ is visible in emission together with a K-star feature at 5168 \r{A}. \citet{2013PASJ...65...23K} found two photometric periods (0.05436 d and 0.05827 d) but the interpretation of them is still not clear. The authors suggested they could be related to superhumps. Spectral observations by \citet{2014MNRAS.441.1186D} reveal CSS1735+15 in eruption with a steep blue continuum and absorption features of H$\gamma$ and H$\delta$.  The SOAR spectrum has the same flux level as the spectrum from \citet{2012AJ....144...81T} and displays typical features of a K star and \ion{Na}{1} absorption together with emission lines of  \ion{H}{1} and \ion{He}{1}. Our spectrum has not enough coverage to allow luminosity class determination, so we cannot discard that the secondary is an evolved object. After the submission of this manuscript, \citet{2016arXiv160902215T} presented time-series spectroscopy and determined an orbital period of 8.48 h and a distance of $1830\pm330$ pc from a K4$\pm$1 spectral type secondary star. This large period challenges the interpretation of the modulations as superhumps, which occurs for systems below the period gap. 
CSS1735+15 is unlikely an mCV.

\subsubsection{1RXS1929+20 (1RXS J192926.6+202038 = USNO-B1.0 1103-0421031)}

This is another CV discovered by \citet{2011AstL...37...91D} among X-ray sources in the ROSAT catalog . It was classified as a possible dwarf nova by its X-ray hardness ratio. Our exploratory spectrum is reddened, with emission lines of \ion{H}{1}, \ion{He}{1} and marginal evidence of \heii, besides two DIBs at 5780 and 6284 \r{A}. The feature at 5577 \r{A} is associated to bad subtraction of the telluric [\ion{O}{1}] line. This CV has no evidence of being a magnetic system.

\subsubsection{MLS2043-19 (MLS100706:204358-194257)}

MLS2043-19 was discovered as a transient optical source by CRTS. Its light curve has a quiescent level at $\sim 19.5$ mag and four registered outbursts, reaching 15 mag. 
The CRTS data do not present any evidence of eclipse or periodicity.
The SOAR spectrum, on the other hand, presents broad and double-peaked Balmer and \ion{He}{1} emission lines, which suggest a high orbital inclination system, besides weak \heii or \ion{C}{3}/\ion{N}{3} emission features over a blue continuum. It also reminds us the spectrum of the IP candidate CTCV2056-30 presented in section~\ref{ctcv}, so we suggest it is a short-period IP candidate.

\subsubsection{MLS2054-19 (MLS101102:205408-194027 = CSS090829:205408-194027)}

This object, discovered by CRTS, has a bimodal light curve with a low-brightness state at 19.5~mag with 1~mag variation and several outbursts reaching 18--16 mag. \citet{2014MNRAS.437..510C} found a photometric period of 0.09598~d, suggesting it to be associated to superhumps of a SU~UMa type CV, and estimated an orbital period of 0.0917 d. Our  exploratory spectrum shows double-peaked H$\beta$ emission more intense than H$\alpha$, besides \ion{He}{1} and \heii (and possibly \ion{He}{2} 4542 \r{A}), being similar to the spectrum of the IP candidate CTCV2056-30, which suggests possible magnetic accretion. Interestingly, if the photometric period was the orbital period, the object would be in the period gap of non-magnetic CVs.

\subsubsection{CTCV2056-30 (CTCV J2056-3014 = 1RXS J205652.1-301433)}
\label{ctcv}

This object was discovered by \citet{2010MNRAS.405..621A} during the Calán-Tololo Survey. Their low-resolution spectra present Balmer and \ion{He}{1} emission lines over a blue continuum. The radial velocity curves provided a period of 1.76~h assumed as the orbital period, while photometric monitoring yields an additional modulation with P=15.4~min, leading to a highly probable IP classification. \citet{2010MNRAS.405..621A} also report some likely dwarf nova outbursts registered by ASAS and the ROSAT X-ray detection as 1RXS J205652.1-301433. Our spectrum shows H$\beta$ as intense as H$\alpha$ and relatively intense \ion{He}{1} and weak \heii lines. The continuum is flatter and about 3 times less intense in the blue than the spectrum from \citet{2010MNRAS.405..621A}. 
The spectrum is quite similar to that of short-period IPs like SDSS~J2333 \citep{2007MNRAS.378..635S,2005AJ....129.2386S}, HT~Cam \citep{2002PASP..114..623K} and DW~Cnc \citep{2004PASP..116..516P,2004MNRAS.349..367R}, among others. These objects present spectra that are indistinguishable from quiescent dwarf novae spectra and, therefore, show that a single optical spectrum is not enough to confirm the IP nature of a source. The definite IP classification demands the detection of the white dwarf spin period not synchronized with the orbital period.

\subsubsection{CSS2108-03 (CSS110513:210846-035031)}

This system was discovered by CRTS in May 2011, displaying a light curve with a quiescent level at 18 mag and outbursts reaching 15 mag. \citet{2014MNRAS.437..510C} performed time-resolved photometry in October 2011 and found eclipses, deriving an orbital period of 0.15699 d. A period of 0.1569268 d was determined independently by \citet{2014MNRAS.441.1186D} from the photometric data in the CRTS itself, in both quiescence and outburst. 

We obtained two exploratory spectra of CSS2108-03, in 2012 May 20 and 2012 June 25. The data display characteristic features of a disk system. Emission lines of \ion{H}{1} and \ion{He}{1} are broad and have pronounced double peak profiles, consistent with the high orbital inclination shown by the eclipses. The continuum flux measured in June was steeper to blue and five times larger than that obtained in May. We do not classify this object as a mCV.

\subsubsection{CSS2200+03 (CSS100624:220031+033431)}

This object was discovered by CRTS and suggested to be a polar type CV in the CRTS circulars\footnote{\url{http://nesssi.cacr.caltech.edu/catalina/Circulars/1006241041184114393.html}}. Its light curve presents a quiescent level with long term (years) variability ranging from 18 to 19.5 mag and several outbursts reaching  about 16.5 mag. \citet{debude} performed photometric follow-up observations, which yield non-eclipsing light curves (within a maximum time span of 3.6 h) dominated by flickering with less than 0.5 mag amplitude. 

Our exploratory spectrum displays broad Balmer and \ion{He}{1} emission lines with double peak profiles, which are more conspicuous in the \ion{He}{1} lines. The presence of a disk, evidenced by the double peak lines and corroborated by the outbursts, discards the polar classification. \heii and \ion{C}{3}/\ion{N}{3} are weak but present. The continuum shows absorption features, which could be associated to a cool secondary. The spectral features are somewhat similar to the spectrum of CTCV2056-30. We classify it as a mCV candidate.

\subsubsection{1RXS2223+07 (1RXS J222335.6+074515 = USNO-B1.0 0977-0743560)}

Also selected as a new CV from the ROSAT catalog, 1RXS2223+07 presents a DN light curve with outbursts 2 mag above the average 19.5 mag quiescent level \citep{2011AstL...37...91D}. Our spectrum presents broad Balmer and \ion{He}{1} emission lines, besides \hheii, on a slightly red continuum, consistent with a DN spectrum.

\subsubsection{SSS2242-66 (SSS110720:224200-662512 = 1RXS J224150.4-662508)}

This source presents a CRTS light curve 
varying between 16.5 and 18~mag with two 
outbursts reaching 15-15.5 mag. Its position matches that of the X-ray source 1RXS J224150.4-662508. Our spectrum, obtained at OPD, shows emission lines of H$\alpha$, H$\beta$, \ion{He}{1} and \heii over a continuum that is slightly steeper to the blue. 
The spectrum is very similar to the spectrum of the IP candidate CTCV2056-30 described above which, as in this case, also presents outbursts and X-ray detection. These similarities may indicate an IP nature for SSS2242-66.

\subsubsection{CSS2319+33 (CSS111021:231909+331540)}

CSS2319+33 has a highly variable CRTS light curve with short and long timescales, ranging from 16.5 to 19.5 mag, but it does not show any outburst. It matches a blue source at SDSS DR8, with $u=17.45, g=17.79, r=17.68, i=17.36$ and $z=17.02$ and also matches the ROSAT source 1RXS~J231909.9+331544. Despite being noisier, the OPD spectrum is very similar to the spectrum of SSS2242-66, which in turn resembles the spectrum of CTCV2056-30, so we consider it as an IP candidate.

\subsection{Systems with Disks in High Accretion State}

\subsubsection{1RXS0721-05 (1RXS J072103.3-055854 = USNO-B1.0 0840-0137592 )}

\citet{2011AstL...37...91D} identified this system as USNO-B1.0 0840-0137592, the optically variable counterpart of an X-ray source in the ROSAT catalog, and suggested a classification as a dwarf nova. Their analysis of archive photographic plates shows that most of the time the system is fainter than 17~mag in the B band, but occasionally gets brighter, reaching 16.5~mag. Our exploratory spectrum has a flux in the V band of $\sim$ 15.3~mag, brighter than in any previous detection, corroborating a dwarf nova classification.
Our spectrum is typical of a dwarf nova in eruption, showing narrow Balmer and \heii emission lines superimposed on a very blue continuum. During DN eruptions, broad absorption troughs may develop at the base of very narrow Balmer emission cores, mainly in the higher member of the Balmer series -- see \citet{warner1995} and the case of SS~Cyg by \citet{1984ApJ...286..747H}. These features are also observed in 1RXS0721-05. Emission lines of \ion{He}{1}, the interstellar absorptions lines of \ion{Na}{1} 5890 and 5896 \r{A}, and the diffuse interstellar band (DIB) at 6284 \r{A} are also present in our spectrum. 
If we consider its ROSAT counts rate \citep{2rxs} and the quiescent optical magnitude, the object is located in the dwarf novae region in Figure 2 of \citet{1993AdSpR..13..115B}. The ROSAT hardness ratios (HR1$ = 1.000 \pm 0.360$ and HR2$ = -0.108 \pm 0.276$) are also consistent with a dwarf nova classification \citep[e.g.,][]{2014OEJV..167....1H}.

\subsubsection{SSS2042-60 (SSS110526:204247-604523)}

SSS2042-60 has a CRTS light curve with two distinct brightness levels at 18 and 19.5~mag, being observed most of the time at the higher state.
No eruption was recorded. Our SOAR spectrum displays a very steep blue continuum with H$\alpha$ in emission and H$\beta$ in broad absorption with a weak central emission feature. \heii is marginally detected and \ion{Na}{1} absorption is visible. Our spectrum has a flux equivalent to V~=~18~mag, near the brightest state. We suggest the object is a novalike system, possibly of VY~Scl type due to the fadings in its light curve.

\subsection{Nova}
\label{sec_nova}

\subsubsection{XMM0630-60 (XMMSL1 J063045.9-603110)}

This source was discovered by XMM-Newton in 2011 as a soft X-ray transient (2011 Dec 01) and was suggested to be a Nova for its X-ray softness \citep{2011ATel.3811....1R}. Observations with GROND identified the optical counterpart with g' = 18.4 mag and r' = 19.5 mag, too faint for a recent Galactic Nova \citep{2011ATel.3813....1K}. The optical source has a blue SED, also atypical for a Nova, with a hint of strong \ion{He}{2} emission. Swift observations show a very soft X-ray spectrum with kT = 48 eV, 12 times fainter than the XMM-Newton flux obtained 19 days before \citep{2011ATel.3821....1R}. XMM0630-60 was also selected as a blazar candidate, but this classification was not confirmed \citep{2013AJ....146..110C}. 

Our spectrum is compatible with a Nova in the nebular phase and displays strong blended lines of [\ion{O}{3}] 4959 and 5007 \r{A}. 
Besides, broad and near rest position emission lines of \hheii, H$\beta$ and H$\alpha$ are also present, with multiple narrow components. Our spectrum corresponds to V $\approx$ 21.8~mag.

\subsection{Miscellaneous Objects}
\label{sec_misc}

In this section we present objects initially considered as mCV candidates due to their CRTS light curves, showing variability that could be interpreted as, for example, distinct high/low states. Many of these candidates have also been pre-classified as CV candidates by CRTS. The exploratory spectra, however, have shown that they are not CVs, but actually can be classified as extragalactic objects -- active galactic nuclei (AGN) which usually share observational features with CVs, like variability and color -- or other types of variable stars as RR Lyrae or low and intermediate-mass pulsating red giants.

The first seven objects described below are classified as extragalactic sources because of the redshift of the lines identified in their spectra. Five of them are dominated by broad \ion{Mg}{2} 2798 \r{A} lines characteristics of AGN. The remaining two objects display both emission and absorption features, with strengths typically observed in galaxies with on-going stellar formation. To the best of our knowledge, we provided here the first classification and redshifts for these sources. The next three objects are RR Lyrae stars and the last three are Mira or semi-regular (SR) pulsating red giants \citep{gcvs1985}.

\subsubsection{CSS0213+00 (CSS091215:021311+002153)}

This CRTS object varies between 20.5 and 18.5 mag and was suggested to be a polar type CV or an AGN\footnote{\url{http://nesssi.cacr.caltech.edu/catalina/Circulars/912151010124106434.html}}. The exploratory spectrum is compatible with a Type~I AGN at z=1.257 based on the conspicuous broad emission line of \ion{Mg}{2} 2798 \r{A} and numerous blended UV \ion{Fe}{2} multiplets. 

\subsubsection{CSS0315-01 (CSS071206:031525-014051)}

CSS0315-01 shows a CRTS light curve slowly varying between V=19 and V=20.5 mag. A broad \ion{Mg}{2} 2798 \r{A} line and the UV \ion{Fe}{2} multiplets in the  spectrum lead to a classification as a Type~I AGN at $z$=0.8513, further confirmed by the detection of [\ion{Ne}{5}]~3425 \r{A}. [\ion{O}{2}]~$\lambda\lambda$3727, 3729 \r{A} are also detected. 

\subsubsection{CSS1240-15 (CSS071218:124027-150558)}

This CRTS transient is highly variable in the range from 18.5 to 21 mag. The spectrum presents narrow emission lines of [\ion{O}{2}] 3726/3728 \r{A}, H$\gamma$, H$\beta$ and [\ion{O}{3}] 4959/5007 \r{A} at a redshift $z=0.272$, besides absorption lines of \ion{Ca}{2} H and K. An upper limit to the flux ratio [\ion{O}{3}]/H$\beta$ is 0.39 , which classifies CSS1240-15 as a \ion{H}{2} galaxy \citep{2006MNRAS.372..961K}, corroborated by the ratio [\ion{O}{2}]/[\ion{O}{3}] = 5.4$\pm$0.6 \citep{1981PASP...93....5B}. The large amplitude variability of this source, on the other hand, leads to a distinct classification as an AGN caught at a minimum state, with the central source completely hidden from direct view , or as a low excitation radio AGN \citep[LERAG,][]{padovani+16}.  Follow-up spectroscopy is necessary to unveil the real nature of this object.

\subsubsection{MLS1251-02 (MLS110329:125118-020208)}

This source is highly variable in CRTS light curve, ranging from 16 to 22 mag in timescales of months. The exploratory spectrum corresponds to an AGN at z=1.1887, determined from a emission line at 6120 \r{A}, which we identify as \ion{Mg}{2} 2798 \r{A}. Absorption doublets are clearly visible at 5700~\r{A} and 5010 \r{A}. These are \ion{Mg}{2} 2796,2803~\r{A} absorption-line systems from two distinct intervening gaseous clouds in the line of sight of the AGN, at redshifts of z$_1=1.037$ and  z$_2=0.791$, respectively.
Absorption features like the doublet of \ion{Fe}{2} 2586,2600 \r{A}, visible at 4640 \r{A}, and the \ion{Mg}{1} 2852 \r{A} line, visible at 5108 \r{A}, are also present and are associated to the gaseous structure  at z$_2=0.791$. The strong variability of MLS1251-02 suggests that it is an optically violently variable (OVV) quasar. 

\subsubsection{MLS1402-09 (MLS110301:140203-090329)}

The CRTS light curve of MLS1402-09 displays a low brightness level at 20.5 average magnitude and a high level at $19.5-20$ mag. The spectrum shows a prominent narrow emission line at 5493 \r{A}, which we identify as [\ion{O}{2}] 3726/3728 \r{A} and from which we derived $z=0.4736$, and absorption lines of \ion{Ca}{2} H and K. In the NVSS catalog there is a radio source (NVSS J140203-090747) offset by 4.307$\arcmin$ from MLS1402-09 with a measured flux of 3.2$\pm$0.6 mJy at 1.4 GHz. This flux level classifies it as a low-emission radio AGN \citep[LERAG,][]{padovani+16}.

\subsubsection{MLS1404-15 (MLS100617:140447-152226)}

MLS1404-15 was suggested to be a CV by CRTS from its blue DSS color and high variability. The CRTS light curve has a low state at 19.5 average magnitude, changing to a high state at average 18.5 mag from 2009 on. A prominent broad feature centered at 5321 \r{A}, that we associated to \ion{Mg}{2} 2798 \r{A} at $z=0.9017$, dominates the spectrum. We also identified the UV \ion{Fe}{2} multiplets at both sides of \ion{Mg}{2} 2798, typical of Type~I AGNs. The presence of high-ionization lines of [\ion{Ne}{5}]~3347,3427~\r{A} and the permitted \ion{O}{3} line at 3134~\r{A} give additional support that MLS1404-15 is indeed an AGN.

\subsubsection{MLS2230-04 (MLS100906:223034-042347)}

Classified as a possible variable star by CRTS, MLS2230-04 has a bimodal light curve with a low state at 20 mag and a high state at 19.3 mag. The SOAR spectrum is dominated by a broad \ion{Mg}{2} 2798 \r{A} emission line and hints of UV \ion{Fe}{2} multiples at both sides of \ion{Mg}{2}, classifying this source as a Type~I AGN  at z=0.968. 

\subsubsection{The RR Lyrae Stars MLS1519-13, MLS2044-16 and MLS2132-15}

\begin{figure}
\plotone{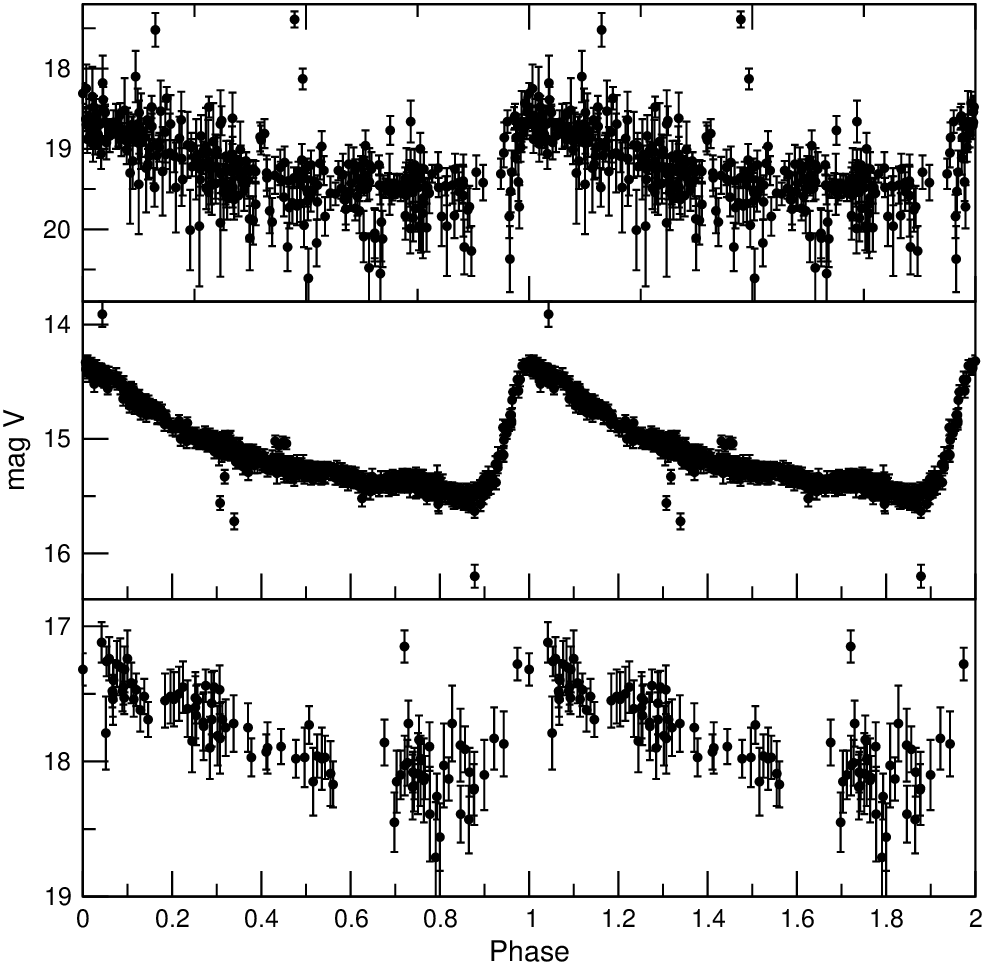}
\caption{CRTS photometric data of MLS1519-13 (top), MLS2044-16 (middle) and MLS2132-15 (bottom) folded with the periods described in the text. \label{figmls1519-13}}
\end{figure}

MLS1519-13, MLS2044-16 and MLS2132-15 were selected for their CRTS light curve variability. The identification spectra are consistent with A or F spectral type, with narrow Balmer absorption lines over a blue continuum. The time-series analysis of the CRTS photometric data yielded periods of 0.521 d for MLS1519-13, 0.535 d for MLS2044-16 and 0.486 d for MLS2132-15. The CRTS light curves of these RR Lyrae stars, folded on the respective periods, are shown in Figure \ref{figmls1519-13}.

\subsubsection{The Pulsating Red Giants Stars SSS1448-40, SSS1639-23 and SSS1940-63}

\begin{figure}
\plotone{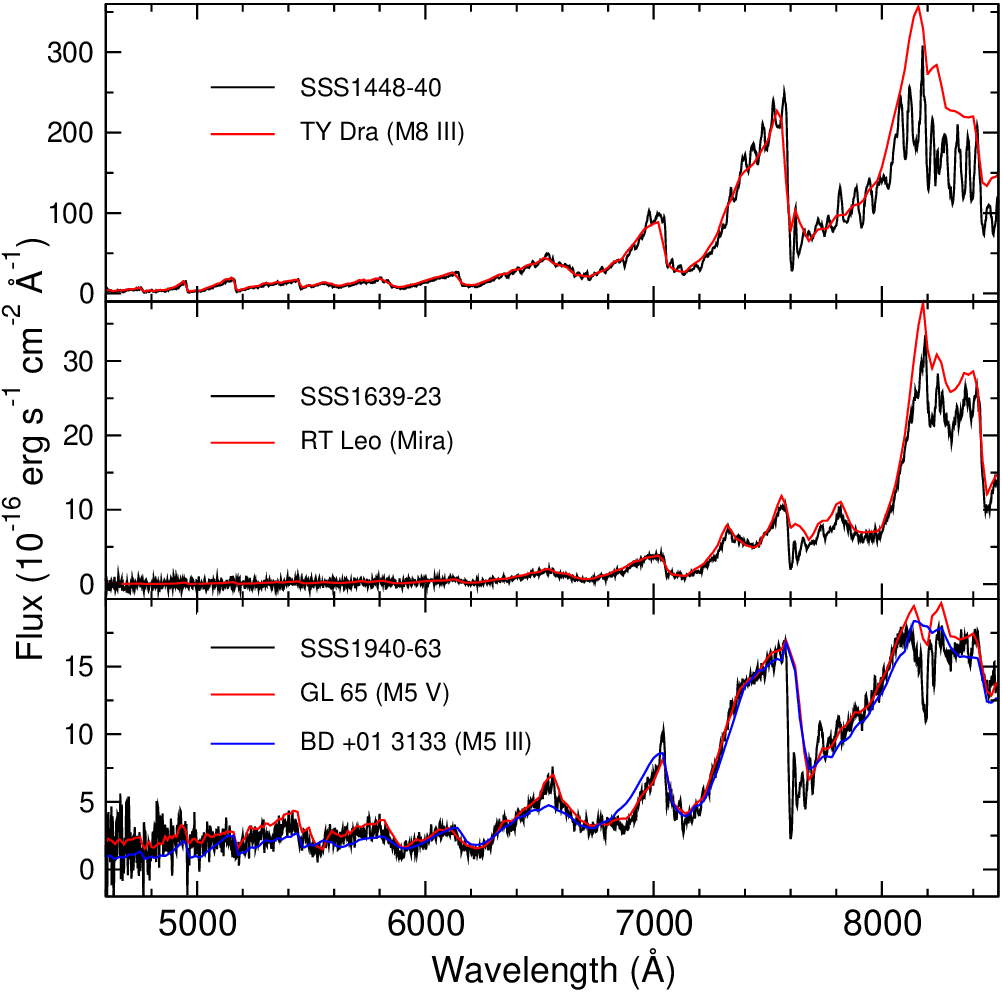}
\caption{Spectra of SSS1448-40 (top), SSS1639-23 (middle) and SSS1940-63 (bottom). The black lines are our spectra and the red and blue lines are the fitted reference spectra used for the spectral type estimation, from the Bruzual-Persson-Gunn-Stryker Atlas \citep{1983ApJS...52..121G}.
} \label{PRGspecs}
\end{figure}

\begin{figure}
\plotone{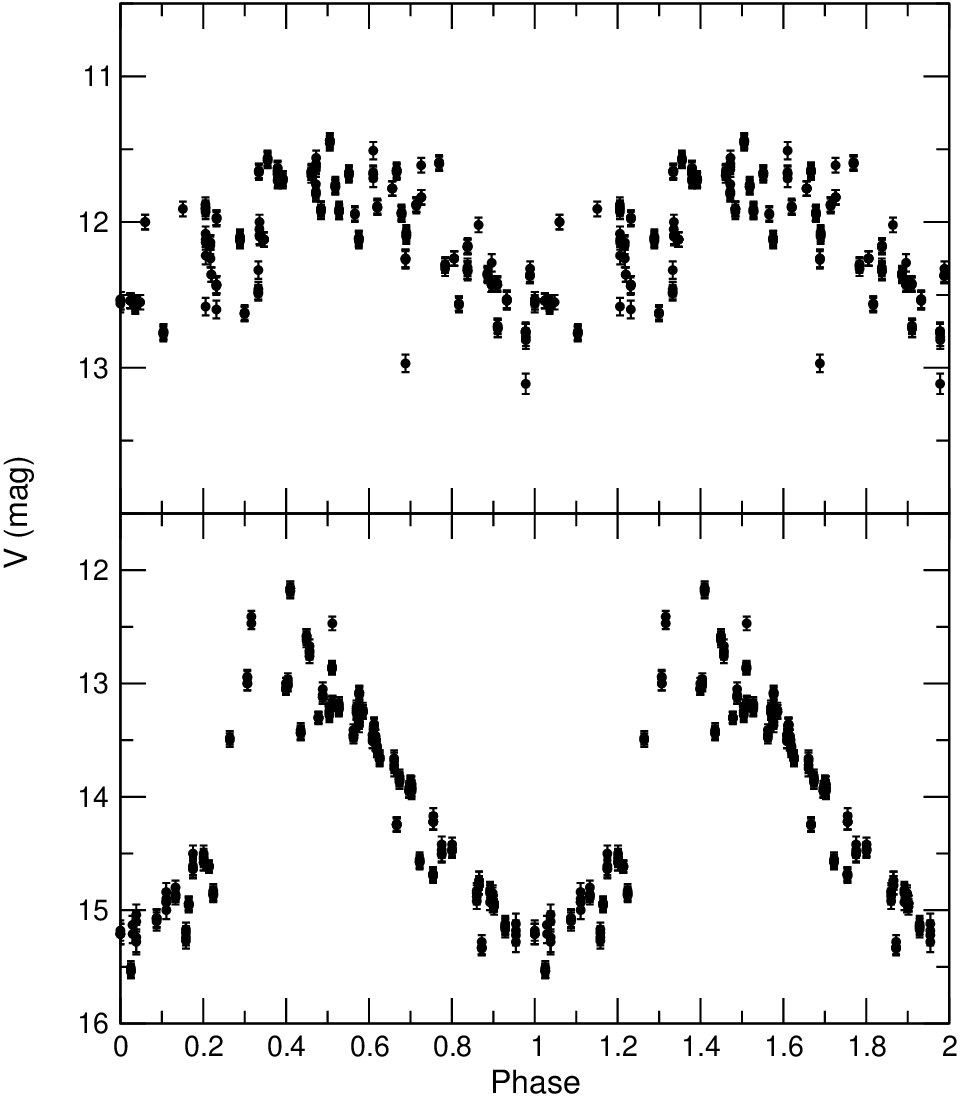}
\caption{CRTS photometric light curves of SSS1448-40 (top) and SSS1639-23 (bottom) folded with periods of 87~d and 307~d, respectively - see text. \label{fig7}}
\end{figure}

SSS1448-40, SSS1639-23 and SSS1940-63 were also selected from the CRTS variability. The exploratory and fitted reference spectra for spectral type estimation are presented in Figure~\ref{PRGspecs}. The DFT and Lomb-Scargle period search methods applied to the CRTS data yielded a period of $87 \pm 1$ d for SSS1448-40 and  $307 \pm 16$ d for SSS1639-23, but no coherent period was obtained for SSS1940-63 (Figure~\ref{fig7}).

\section{SUMMARY AND DISCUSSION}

We report on an exploratory spectroscopy survey of magnetic CVs based on objects selected mainly from the CRTS photometric transient survey but also from the literature. The main purpose of this work is to improve the statistics of the mCV class. From an initial sample of 45 objects, we found 32 sources which we classified as CVs based on the aspect of the exploratory spectra. This group is composed of 13 polar candidates, 9 IP candidates, 7 DNs in quiescence, 2 objects in high-accretion state and 1 nova. In addition to CVs, we have also found RR~Lyrae stars, variable red giants and extragalactic objects. These results are summarized below. 

From the group of 13 polar candidates, 5 are new classifications (MLS0456+18, MLS0720+17, MLS0854+13, CSS1127-05 and MLS1609-10) and other 6 are already suggested polars for which we provide the first spectrum and confirm the classification. The remaining 2 polar candidates present notable spectral variability relative to published spectra.  
In the \citet{2003A&A...404..301R} catalog (update RKcat7.23, 2015), there are 119 confirmed and 30 uncertain polars. Six objects of our sample are not included in RKcat7.23. Only 6 objects (out of 13 polar candidates) are known high-energy sources, which shows that the X-rays detection is not a {\it sine qua non} criterion for the classification as a polar 
(or as a magnetic CV, see also \citet{2014ASPC..490..389M}). This is particularly true for faint/distant objects, which could be missed by soft X-rays surveys - see, for instance, the discussion of MLS0227+13 presented in \citet{2015MNRAS.451.4183S}. Ten objects obey the two Silber's criteria for mCV classification \citep{silber1986}, hence Silber's criteria seems to be adequate to define polars. The three remaining objects are peculiar in some way: 1RXS1002-19, which has emission lines too broad for a polar; CSS1127-05, which has the largest H$\beta$ EW of our CV sample; and CSS1503-22, which presents absorption features in the H$\beta$ region preventing a proper quantification of line indicators. We present the first estimates for the orbital periods of MLS0456+18 (P~=~2.36~h) and MLS1609-10 (P~=~1.85~h).

We classify 16 objects as probable CV with quiescent disks, including 1 confirmed IP and 8 possible IPs. CTCV2056-30 is a strong IP candidate due to its photometric variability modulated by the white dwarf rotation \citep{2010MNRAS.405..621A}. Most of the IP candidates present the \heii line, although in low intensity. CSS0104-03,  however, has this line strong enough to have its EW measured and, even in this case, its EW is smaller than 40\% of H$\beta$ EW, therefore too weak to obey the Silber's criteria for a mCV. Even so, we classify it as a probable IP because its \heii line is stronger in quiescence compared to the brighter state, contrary to what is expected from non-magnetic CVs. The 7 remaining new IP candidates (namely CSS1012-18, SSS1359-39, MLS2043-19, MLS2054-19, CSS2200+03, SSS2242-66 and CSS2319+33) were classified as such by spectral similarity to CTCV2056-30. All our nine IP candidates have \heii in the spectra -- but none obeys Silber's criteria -- 4 of them have X-ray counterparts and 8 have shown outbursts.

The spectrum of CTCV2056-30, whose orbital period is around 1.75~h, is very similar to the spectrum of several short-period IPs, like SDSS J2333, HT~Cam and DW~Cnc. Could our IP candidates spectrally similar to CTCV2056-30 have short periods? While most of the polars have orbital periods below 4 h, few IPs have periods below the 2--3 h period gap. The \citet{2003A&A...404..301R} catalog registers 68 confirmed IPs and only 9 of those have orbital periods below the gap. This was interpreted as an evolution of long-period/high-B IPs to short period polars, whereas those with lower magnetic fields remain as IPs \citep{2004ApJ...614..349N}. This kind of conclusion demands a sample of short period IPs as complete as possible. Follow-up studies of these candidates are essential, and may confirm their nature and help to build an homogeneous sample of IPs, which can be used to investigate the issue of mCV evolution. 

Ten remaining objects from our sample are CVs as well. Seven are DNs with quiescent disks and two have disks in high state of accretion: the known dwarf nova 1RXS0721-05, which we caught in eruption, and the newly discovered CV SSS2042-60, which we classify as a novalike variable, possibly of VY~Scl type. We also confirmed that the transient high-energy source XMM0630-60 is a nova, which we observed in the nebular phase.

Previous similar works \citep[e.g.,][]{2014MNRAS.441.1186D,2011AJ....142..181S,2014AJ....148...63S,2012AJ....144...81T,2016arXiv160902215T} targeted on the discovery of CVs from variability or color-based selected samples, have indirectly shown that the identification of a mCV is not a trivial task. Usually, no isolated observational technique is enough to confirm the classification as a mCV, specially in the case of IPs. The usual criteria include X-ray detection, the relative intensity of the \ion{He}{2} 4686~\r{A} emission line and polarized optical/IR emission. But none of these criteria is essential to the mCV classification, while some are not exclusive of the mCV class. A mCV may not be detected as an X-ray source and, on the other hand, many DNs are X-ray emitters. The high-ionization line of \ion{He}{2} 4686~\r{A} is present, for instance, in non-magnetic CVs like post-novae, novalike systems and close binary supersoft X-ray sources \citep[CBSS,][]{2006AdSpR..38.2836K,2004MNRAS.351..685O} and, with lower intensity, in DNs as well. Also, few IPs have polarized emission. However, variability-based selection followed by spectroscopic snapshot observations of the selected targets may be a first step for the discovery of mCVs, being a relatively inexpensive strategy in terms of telescope time. More than 70$\%$ of our sample are CVs and almost 50$\%$ of the sample was classified as mCVs. If the classifications from this work are confirmed, it would represent an increase of about 4$\%$ in the number of known polars and $12\%$ in the number of known IPs. The confirmation of the classifications and the detailed nature of the CVs will be determined by time-resolved spectroscopic, photometric and polarimetric follow-up observations in the future. 

\vspace{5mm}

\acknowledgments
We would like to thank the anonymous referee for useful comments, which helped to improve the paper. 
We also thanks J. E. Steiner and M. P. Diaz for important suggestions on the classification of some objects. This study was partially supported by CNPq (CVR: 306701/2015-4 and 306103/2012-5, KMGS:302071/2013-0, ARA: 311935/2015-0), FAPESP (LAA: 2012/09716-6 and 2013/18245-0; CVR: 2013/26258-4) and CAPES (MSP: 23038.009634/2016-71). This work is based partly on observations made at the Observatório do Pico dos Dias, Brazil, operated by the Laboratório Nacional de Astrofísica,
and at the Southern Astrophysical Research (SOAR) telescope, which is a joint project of the Ministério da Ciência, Tecnologia e Inovação (MCTI) da República Federativa do Brasil, the US National Optical Astronomy Observatory
(NOAO), the University of North Carolina at Chapel Hill (UNC) and Michigan State University (MSU).
The CRTS survey is supported by the US National Science Foundation under grants AST-0909182.
The CSS survey is funded by the National Aeronautics and Space
Administration under grant no. NNG05GF22G issued through the
Science Mission Directorate Near-Earth Objects Observations Program.
STSDAS is a product of the Space Telescope Science Institute, which is operated by AURA for NASA. 

\vspace{5mm}

\facilities{SOAR (Goodman HTS), LNA:1.6m}

\software{IRAF}

\end{document}